\documentclass[smallextended]{svjour3}       
\smartqed  
\usepackage{graphicx}
\usepackage{euscript,amsmath,amssymb,amsfonts,graphicx,fullpage,bm,color,cite,paralist}

\usepackage{natbib}

\numberwithin{equation}{section}

\usepackage{epstopdf} 

\newcommand{\e}{{\rm e}}
\renewcommand{\d}{{\rm d}}

\newcommand{\D}{\displaystyle}
\newcommand{\mc}{\mathcal }

\begin{document}

\title{Sensory feedback in a bump attractor model of path integration}

\titlerunning{Certainty in bumps}        

\author{Daniel B. Poll \and Khanh Nguyen \and Zachary P. Kilpatrick}

\institute{D.B. Poll \at
              Department of Mathematics, University of Houston, \\
              Houston TX 77204 USA \\
              \email{dbpoll@math.uh.edu} 
              \and
              K. Nguyen \at
              Department of Mathematics, University of Houston, \\
              Houston TX 77204 USA \\
              \email{kpnguyen21@yahoo.com}
              \and
              Z.P. Kilpatrick \at
              Department of Mathematics, University of Houston, \\
              Houston TX 77204 USA \\
              \email{zpkilpat@math.uh.edu}    
}

\date{Received: date / Accepted: date}

\maketitle

\begin{abstract}
The mammalian spatial navigation system makes use of several different sensory information channels. This information is then converted into a neural code that represents the animal's current position in space by engaging place cell, grid cell, and head direction cell networks. In particular, sensory landmark (allothetic) cues can be utilized in concert with an animal's knowledge of its own velocity (idiothetic) cues to generate a more accurate representation of position than (idiothetic) path integration provides on its own \citep{battaglia04}. We develop a computational model that merges path integration with information from external sensory cues that provide a reliable representation of spatial position along an annular track. Starting with a continuous bump attractor model, we allow for the possibility of synaptic spatial heterogeneity that would break the translation symmetry of space. We use asymptotic analysis to reduce the bump attractor model to a single scalar equation whose potential represents the impact of heterogeneity. Such heterogeneity causes errors to build up when the network performs path integration, but these errors can be corrected by an external control signal representing the effects of sensory cues. We demonstrate that there is an optimal strength and decay rate of the control signal when cues are placed both periodically and randomly. A similar analysis is performed when errors in path integration arise from dynamic noise fluctuations. Again, there is an optimal strength and decay of discrete control that minimizes the path integration error.

\keywords{neural field, sensory feedback, spatial navigation, stochastic differential equation}
\end{abstract}

\section{Introduction}
\label{intro}

\begin{figure}[t]
\begin{center} \includegraphics[width=7.8cm]{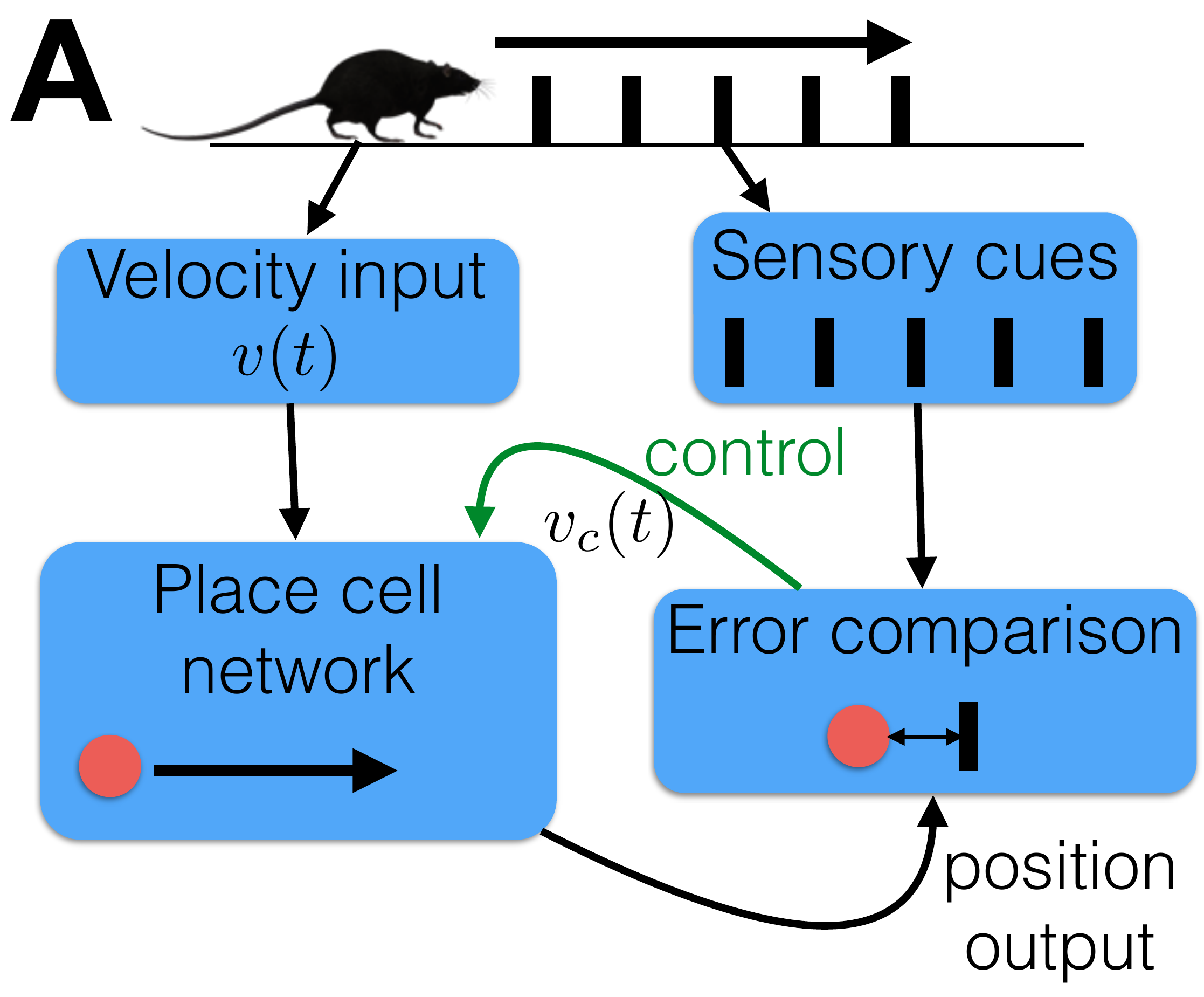} \includegraphics[width=6.6cm]{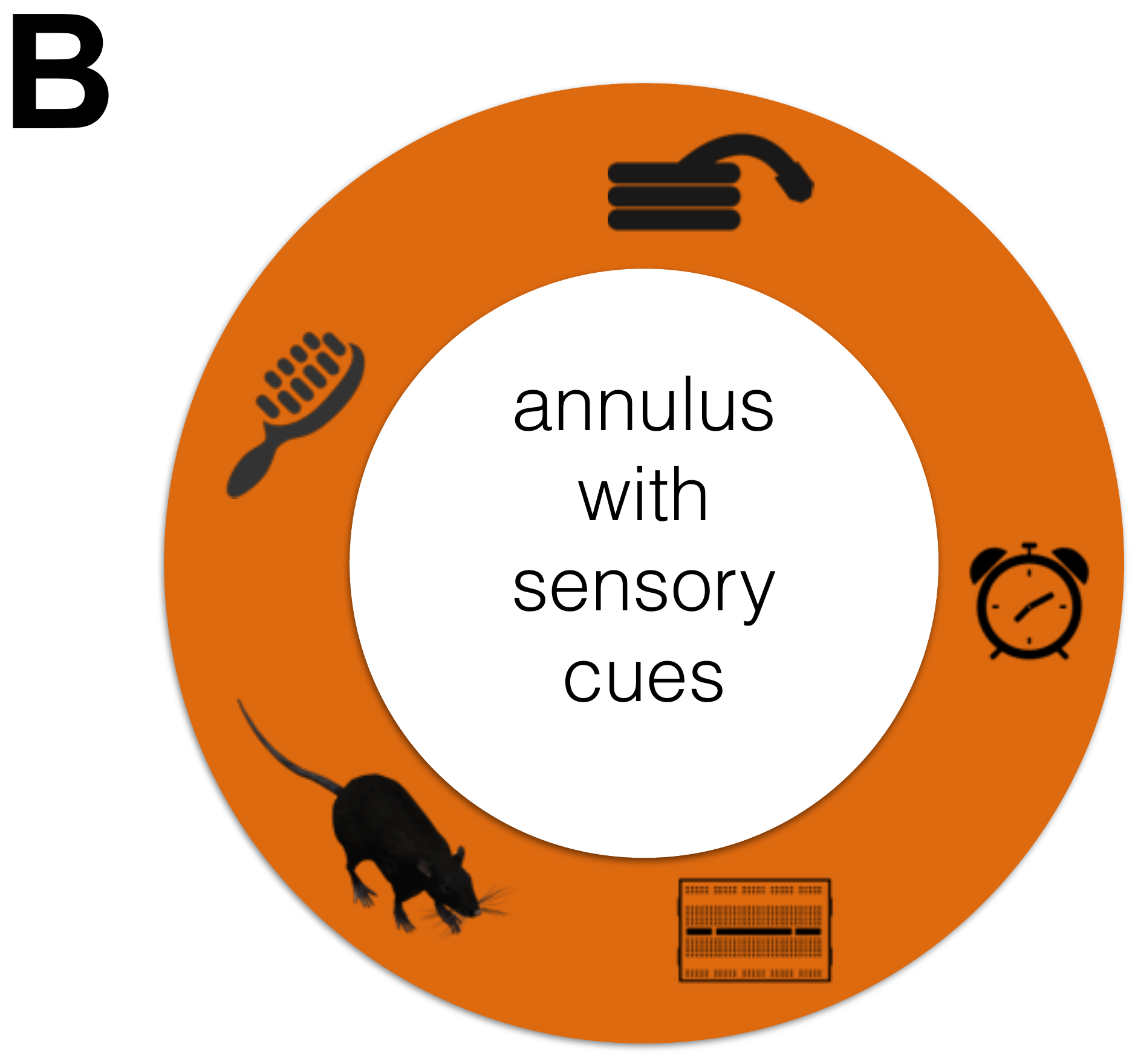} \end{center}
\caption{Mammalian spatial navigation network with sensory feedback. The animal utilizes its own velocity to update its remembered position (path integration) and corrects this memory with sensory cues that serve as position landmarks. ({\bf A}) Schematic of the underlying neuronal network demonstrates the place cell network receives direction input from the animal's velocity signal $v(t)$. Its position estimate is compared with the true position read-out from a sensory cue, and this error is then used to generate a control input signal $v_c(t)$ back into the place cell network. ({\bf B}) Illustration of the experiments by \cite{battaglia04}, showing an annular track with various object placed in the environment to provide the animal with sensory cues.}
\label{fig1}
\end{figure}

Animals have a remarkable ability to accurately navigate over large distances \citep{geva15}. For instance, birds can utilize neural systems that sense the earth's magnetic field, orienting themselves geocentrically \citep{cochran04,wu12}. This is in contrast to the systems studied in mammalian species, who are typically shown to use path integration \citep{etienne96,mcnaughton06}. Path integration models of spatial navigation assume mammals have knowledge of their direction and speed of motion, which networks of the brain can then integrate to encode the path of their idiothetic motion \citep{samsonovich97}. However, there are a number of potential sources of error to this mechanism. The nervous system itself is prone to a wide variety of noise sources due to channel fluctuations, synaptic failures, or even stochastic network-wide events \citep{faisal08}. This could lead to faulty communication of velocity or head direction signals, or it could corrupt the storage of the position signal. Furthermore, the network that integrates the velocity signal may be comprised of architecture that is heterogeneous, providing an imperfect summation of velocity inputs \citep{brody03}. Any inaccuracy in the represented position or velocity will be compounded over time, as the neural code continues to trace the animal's true path \citep{knierim95,valerio12}. Fortunately, path integration is not the sole navigational technique of the mammalian brain; landmarks detected by the sensory system help anchor and correct the integrated velocity signal (Fig. \ref{fig1}{\bf A}) \citep{etienne96,collett04,solstad08}.

 Several experiments have demonstrated that mammals' representation of space is sharpened in the presence of sensory cues \citep{battaglia04,ulanovsky11,aikath14,zhang14}. Experiments typically compare place fields of individuals cells - spatial locations where the cell becomes active - in the presence and absence of sensory landmarks (e.g., steel brush or ticking clock; Fig. \ref{fig1}{\bf B}). For instance, \cite{battaglia04} recorded from hippocampal place cells in rats moving on an annular track. When there were no sensory cues along the track, the place fields of individual cells differed substantially, depending on whether the rat was moving clockwise or counterclockwise around the annulus. This suggests there was some drift in animals' neural representation of their position. However, when several position landmarking cues were placed along the track, the clockwise and counterclockwise place fields of individual cells were considerably correlated. This suggests the sensory cues tightened the navigation systems fine representation of the animal's spatial position. Similar effects have been observed in brown bats, whose echolocation signals provide a brief burst of rich sensory information, sharpening the animal's place fields \citep{ulanovsky11}. Thus, sensory cues appear to provide a correction mechanism for the many sources of error that disrupt position representation and broaden place fields.

We bring these two features of the mammalian navigation system -- path integration and sensory feedback -- together into a single model. Our main focus is the role sensory feedback can play in correcting the path integration signal. Errors in the path integration signal will arise, in our model, due to internal disruptions of an accurately delivered velocity input. The model represents position by utilizing a perturbation of a continuous bump attractor \citep{amari77,zhang96}. Bump states arise in these models due to a combination of strong local excitation and broadly tuned inhibitory feedback \citep{wang99}. In translation symmetric networks, bumps can be formed with their center of mass at any location. However, such well-balanced architecture is unlikely to occur in actual networks of the brain, which tend to be spatially heterogeneous \citep{brody03,renart03}. Such symmetry-breaking in the bump attractor network leads to system states that drift toward a finite number of discrete attractors, so the long term dynamics are weakly correlated with the input signal \citep{zhang96,itskov11,kilpatrick13b}. Furthermore, any dynamic fluctuations in the voltage or synaptic signals of the network can lead to diffusive wandering of the bump state that will also degrade its signal representation \citep{compte00,burak12,kilpatrick13}. Our study mainly focuses on how external control can reduce the deleterious effects of both spatial heterogeneity and noise, ultimately improving long term accuracy of the network's path integration.

Our model is intended to describe the neural activity of place cell networks in the hippocampus \citep{okeefe96}. Based on a simplified version of the model by \cite{samsonovich97}, neural activity bumps are propelled around the network by external velocity inputs that introduce spatial asymmetry into the balance of excitation and inhibition. A similar mechanism was utilized by \cite{burak09} in a two-dimensional model of grid cell activity. Thus, this framework is a well accepted model of position encoding cells in hippocampus, entorhinal cortex, and the vestibular system \citep{zhang96}. Since we are modeling motion along an annulus (Fig. \ref{fig1}{\bf B}), we have restricted the network to a one-dimensional periodic domain. Sensory cues are assumed to provide a reliable estimate of the animal's true current position. This position is then compared with the place cell network's estimate of position. Any discrepancy in the position estimate is then translated into a corrective velocity input, which is added to the baseline velocity input (Fig. \ref{fig1}{\bf A}). Even when the cues occur discretely in space, this mechanism works well for reducing the long term error in the position estimate.

In section \ref{model}, we introduce the neural field model of spatial navigation, which combines path integration and sensory cue feedback. Next, we derive a low-dimensional approximation for the dynamics of bump position in the neural field model (section \ref{potwell}). This reduction reveals the relative influence of velocity inputs, sensory feedback, and heterogeneity on the animal's perceived position of its current location. Ultimately, this allows us to calculate the impact of various control strategies on the error between the animal's perceived position and true position (section \ref{control}). Our main finding is that there is an optimal control strength at which the long term error of the network is minimized. Our findings were similar in the case that errors arose due to dynamic noise fluctuations (subsection \ref{addnoise}), rather than synaptic heterogeneities (subsection \ref{reducehet}). In this case, the low-dimensional approximation of the neural field is a stochastic differential equation whose variance we can evaluate explicitly.

\section{Sensory control in velocity-integrating place cell networks}
\label{model}

We employ a neural field model of velocity integration that sustains a bump attractor of neural activity in the absence of any inputs. \cite{amari77} pioneered the scalar neural field model as a reduction of the excitatory-inhibitory model of \cite{wilson73}, but the incorporation of velocity inputs that shift the bump around the spatial domain is more recent. Originally developed as a model of the head direction system \citep{zhang96}, velocity-integrating networks introduce an external input that alters the shape of the recurrent architecture \citep{mcnaughton91}. As a result, a moving bump, rather than a stationary bump, becomes the stable solution to the model equations. This model has since been extended to account for place fields and grid cell fields in planar systems \citep{samsonovich97,burak09}. The fully general form of our neural field model is given
\begin{equation}
d u (x,t) = - \left[ u(x,t) +  \int_{-\pi}^\pi w(x,y) f(u(y,t)) dy + \tilde{v}(t) \int_{-\pi}^\pi w_v (x-y) f(u(y,t)) dy \right] dt + \epsilon dW(x,t) \label{nfmodel}
\end{equation}
where $u(x,t)$ denotes the total synaptic activity at a position $x \in [-\pi,\pi]$  at some point in time $t$. While $x$ labels the position of neurons in the network it also corresponds to location in the environment, so the domain $\Omega = [- \pi, \pi]$ is taken to be periodic as it represents an annular track (Fig. \ref{fig1}{\bf B}).

The function $w(x,y)$ represents the synaptic connectivity between neurons, which we model as a translationally symmetric unimodal function $w_0$, modified by spatial heterogeneity $w_u$ or odd asymmetry $\phi$, so
\begin{equation}
w(x,y) := (1 + \sigma w_u (y)) w_0 (x-y - \phi) \label{wf}
\end{equation}
with $\sigma, \; \phi \ll 1$. Note that in the limit $\sigma \to 0$ and $\phi \to 0$, we obtain $w(x,y) = w_0(x-y)$, a distance-dependent even function. However, in the fully general case ($\sigma > 0$ and/or $\phi>0$), it is straightforward to see that the function $w(x,y)$ need not be distance-dependent or even symmetric. In particular, when $\sigma >0$, discrete attractors form in the network (\ref{nfmodel}) whereby bumps tend to drift away from their initial position to a finite number of linearly stable locations \citep{zhang96,itskov11,kilpatrick13}. We consider this to be a major source of error in the network, since near-perfect integration of the velocity inputs could be achieved if $w(x,y) = w_0(x-y)$. For ease of analysis, the translationally symmetric function is typically taken to be a cosine $w_0(x) := \cos x$. We will allow $w_u$ to be arbitrary, by representing it as a series of $N$ Fourier modes
\begin{align}
w_u(x) := \sum_{n=1}^N \alpha_n \cos(nx) + \beta_n \sin(nx); \hspace{3mm} \langle \alpha_n \rangle = \langle \beta_n \rangle = 0;  \hspace{3mm} \langle \alpha_n^2 \rangle = \langle \beta_n^2 \rangle = \sigma_n^2.  \label{ewf} 
\end{align}
The coefficients $\alpha_n, \beta_n$ are random variables drawn from the normal distribution with mean zero and variance $\sigma_n$. 

\begin{figure}[t]
\begin{center} \includegraphics[width=5cm]{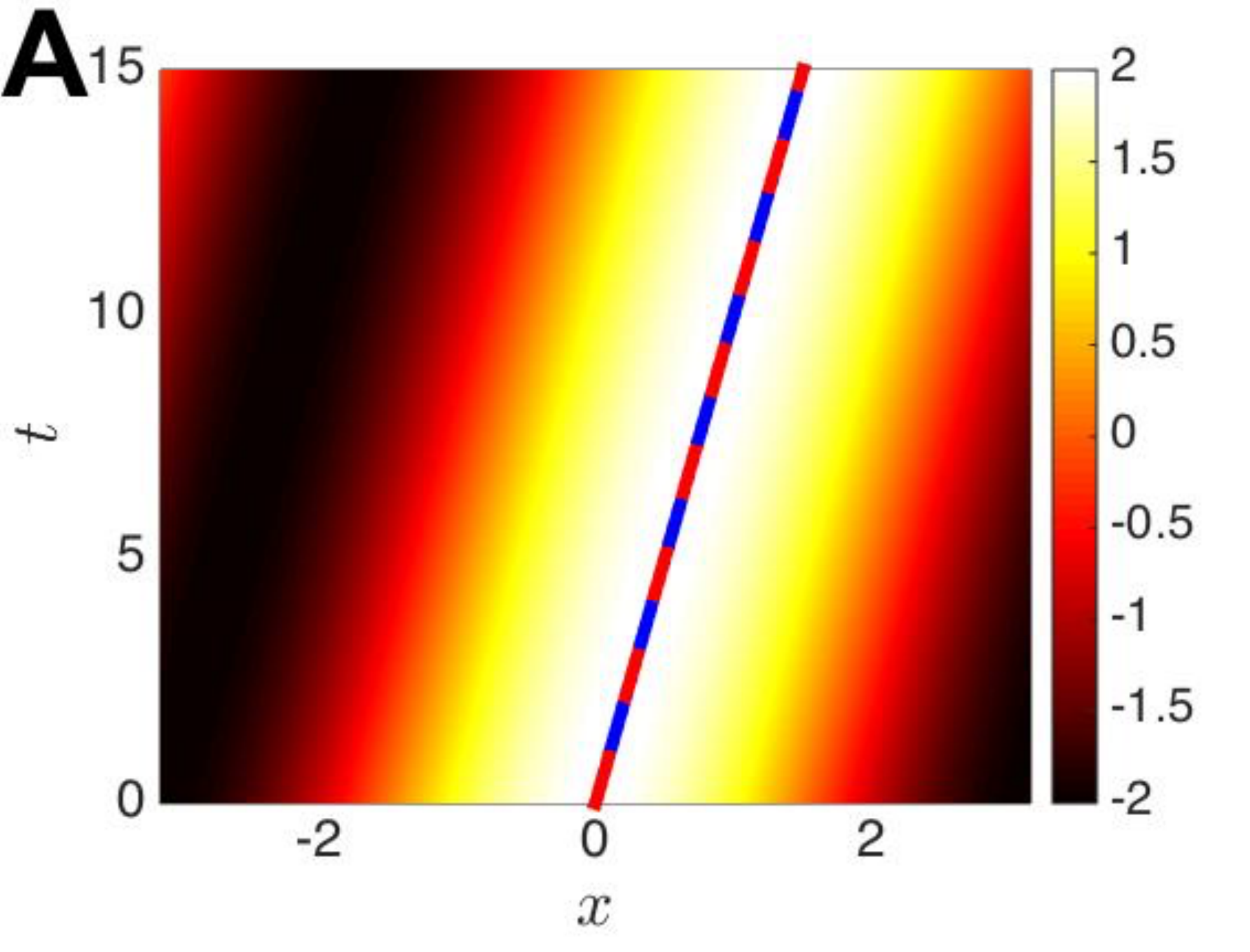}
\includegraphics[width=5cm]{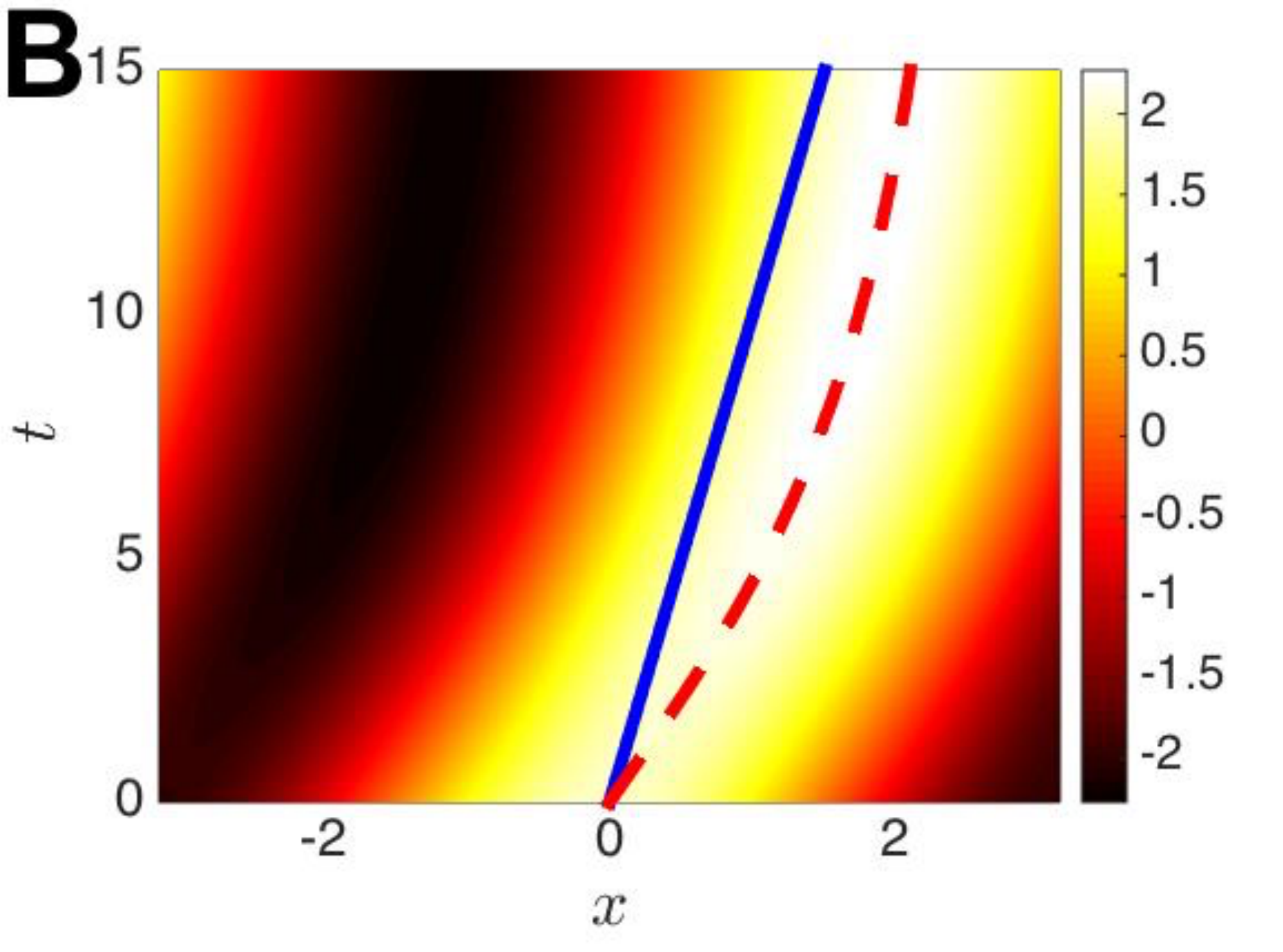}
\includegraphics[width=4.9cm]{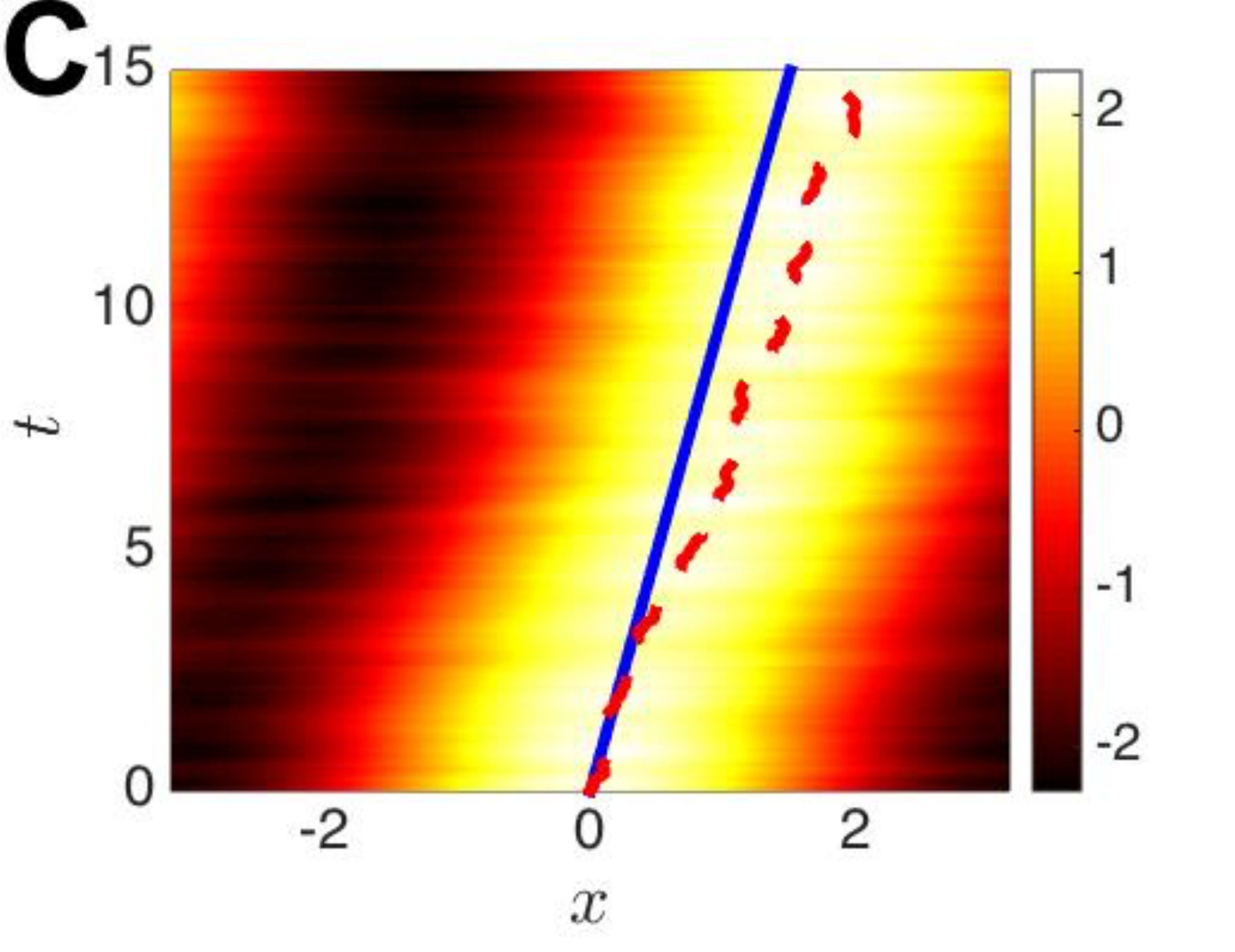} \end{center}
\caption{Velocity-integration with moving bump attractors in the neural field model (\ref{nfmodel}) with a Heaviside firing rate function (\ref{H}) with threshold $\theta = 0$ and a cosine base weight function $w_0(x) = \cos (x)$. ({\bf A}) Bump of neural activity $u(x,t)$ perfectly integrates velocity inputs in the case of no heterogeneity ($\sigma = 0$) and no noise ($\epsilon = 0$), showing the animal's true position (solid line) is perfectly tracked by the center of mass of the bump (dashed). ({\bf B}) In a heterogeneous network ($\sigma = 0.1$ with $w_u =  \sin (x)$), the bump initially moves too fast due to a discrete attractor of the input-free system at $x = \pi/2$, so the bump's center of mass is mismatched with the true position of the animal. ({\bf C}) In the presence of spatiotemporal noise ($\epsilon = 0.2$ with cosine correlations $C(x) = \cos (x)$), the bump wanders diffusively so the encoded position tends to slowly distance itself from the true position.  Here the external velocity input is constant $\tilde{v}(t) = 0.1$. Numerical simulations are performed using an Euler timestep with $dt = 0.1$ and a trapezoidal rule on the integral with $dx \approx 0.003$.}
\label{fig2}
\end{figure}

Velocity inputs are represented by the shifting function $w_v(x-y) := -w_0'(x-y)$ as in the original head direction system model \citep{zhang96} and recent grid cell models \citep{burak09}. In the absence of any heterogeneity or asymmetry, the sum $w(x,y) + \tilde{v}(t) \cdot w_v(x-y)$ would be translation symmetric but not even symmetric in general. This asymmetry produces a moving bump as the solution to (\ref{nfmodel}) that will move at a speed given by $|\tilde{v}(t)|$ (Fig \ref{fig2}{\bf A}). Incorporating heterogeneity, $\sigma >0$, the system is no longer translation symmetric, and a moving bump will not move at the same speed as the velocity input $|\tilde{v}(t)|$ (Fig. \ref{fig2}{\bf B}). Thus, assuming a sensory mechanism for correcting the place cell's encoded position when a cue is encountered, take the velocity input to be
\begin{align*}
\tilde{v}(t) := v(t) + v_c(t),
\end{align*}
the sum of the animal's true velocity $v(t)$ and an external control signal $v_c(t)$. This is meant to account for the improved place representation observed when animals can employ information about sensory landmarks \citep{battaglia04,ulanovsky11,aikath14}. As shown in the schematic in Fig. \ref{fig1}{\bf A}, we assume there is a network that can access the place cell network's perceived position $\Delta (t)$ via a readout of the center of mass of neural activity \citep{deneve99}
\begin{align}
\Delta (t) = \int_{- \pi}^{\pi} x f(u(x,t)) \d x. \label{combump}
\end{align}
The present positional error is then computed by comparing the perceived position $\Delta (t)$ to the animal's actual position given by a time integral of the velocity input
\begin{align*}
\Delta_T(t) = \int_0^t v(s) \d s,
\end{align*}
so the error
\begin{align}
r (t) = \Delta_T(t) - \Delta (t), \label{err}
\end{align}
which will be positive (negative) if the estimated position is to the left (right) of the true position. Note, we extend the domain $x \in [- \pi, \pi]$ to compute (\ref{err}) in cases where the closest distance between $\Delta_T$ and $\Delta$ is across the boundary cuts at $x = \pm \pi$. The error $r(t)$ is then translated either into a continuous velocity control signal
\begin{align}
v_c(t) = \lambda r(t) = \lambda \cdot ( \Delta_T(t) - \Delta (t)), \label{contcontr}
\end{align}
or a discrete control signal given by
\begin{align}
\frac{\d v_c}{\d t} = - \frac{v_c(t)}{\tau} + \lambda \sum_{k=1}^{N_c} r (t_k) \delta (t-t_k),  \label{disccontr}
\end{align}
where sensory cues occur at times $t_k$ and $\lambda$ and $\tau$ determine the strength and time decay of control. As we will show, in the case of continuous control (\ref{contcontr}) strengthening the sensory feedback $\lambda$ always leads to a reduction of the error. This is not the case for discrete control (\ref{disccontr}), since the previous sensory cue at $t_k<t$ becomes less relevant as $t$ increases toward $t_{k+1}$. One of the main goals of this study is to explore how the spacing between subsequent cues $t_{k+1} - t_k$ determines how strong $\lambda$ the control signal should be.

The nonlinearity $f$ is a firing rate function taken to be sigmoidal \citep{wilson73}
\begin{equation*}
f(u) := \frac{1}{1 + e^{-\gamma(u - \theta)}}
\end{equation*}
where $\gamma$ is the gain and $\theta$ is the firing threshold. For ease in analysis, we will often consider the high gain limit $\gamma \rightarrow \infty$ so that $f$ becomes a Heaviside step function of the form
\begin{equation}
f(u) := H(u - \kappa) = \begin{cases} \label{H}
                         1 : &u \geq \theta,  \\
                         0 : &\text{ otherwise}.
                        \end{cases}
\end{equation}

Lastly, we also will consider the impact of the additive noise increment $dW(x,t)$. Spatially extended Langevin equations of the form (\ref{nfmodel}) have become a common model of the effects of fluctuations in large-scale neuronal networks \citep{bressloff12}. The noise term is a spatially filtered spatiotemporal white noise process
\begin{align*}
dW(x,t) := \int_{-\pi}^\pi \mathcal{F}(x-y) \; dY(y,t)  dy,
\end{align*}
where $\mathcal{F}$ is the spatial filter and $dY(x,t)$ is a spatially and temporally white noise increment. With these definitions, the mean and variance can be calculated as $\langle dW(x,t) \rangle = 0$ and $\langle dW(x,t) \; dW(y,s) \rangle = C(x-y) \delta(t-s)$, where $\delta (t)$ is the delta function and  $C$ is the spatial correlation given in terms of the filter $\mathcal{F}$ as
$$
C(x-y) = \int_{-\pi}^\pi \mathcal{F}(x - x') \mathcal{F}(y - x') dx'.
$$
As an example, consider $\mathcal{F}(x) = \cos(x) + \sin(x)$. Then, the spatial correlation $C$ can be computed explicitly as
\begin{align*}
&\int_{-\pi}^\pi \mathcal{F}(x - x') \mathcal{F}(y - x') dx' = \int_{-\pi}^\pi  \big( \cos(x - x') + \sin(x - x') \big)   \big( \cos(y - x') + \sin(y - x') \big) dx' \\
 =  \; &\int_{-\pi}^\pi \cos^2(x') \cos(x) \cos(y) + \sin^2(x') \sin(x) \sin(y) dx' = \pi \big( \cos(x) \cos(y) + \sin(x) \sin(y) \big) \\
=  \; & \pi \cos(x-y) =: C(x-y).
 \end{align*}
 As we demonstrate, the control introduced to account for the impact of synaptic spatial heterogeneity can also be utilized to decrease errors brought about by spatiotemporal noise (Fig. \ref{fig2}{\bf C}).
 
 \begin{figure}[t]
 \begin{center}
\includegraphics[width=5cm]{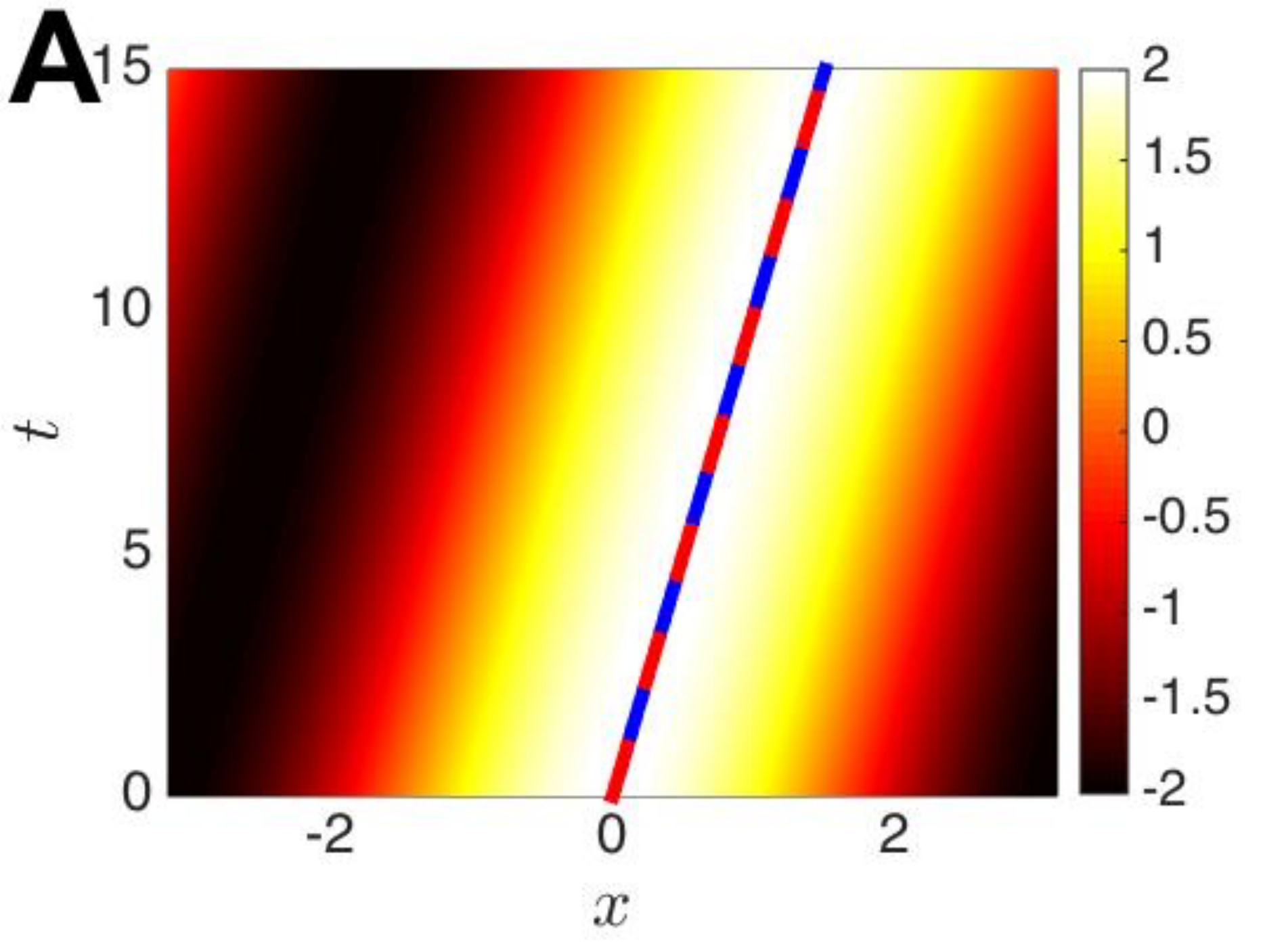}
\includegraphics[width=5cm]{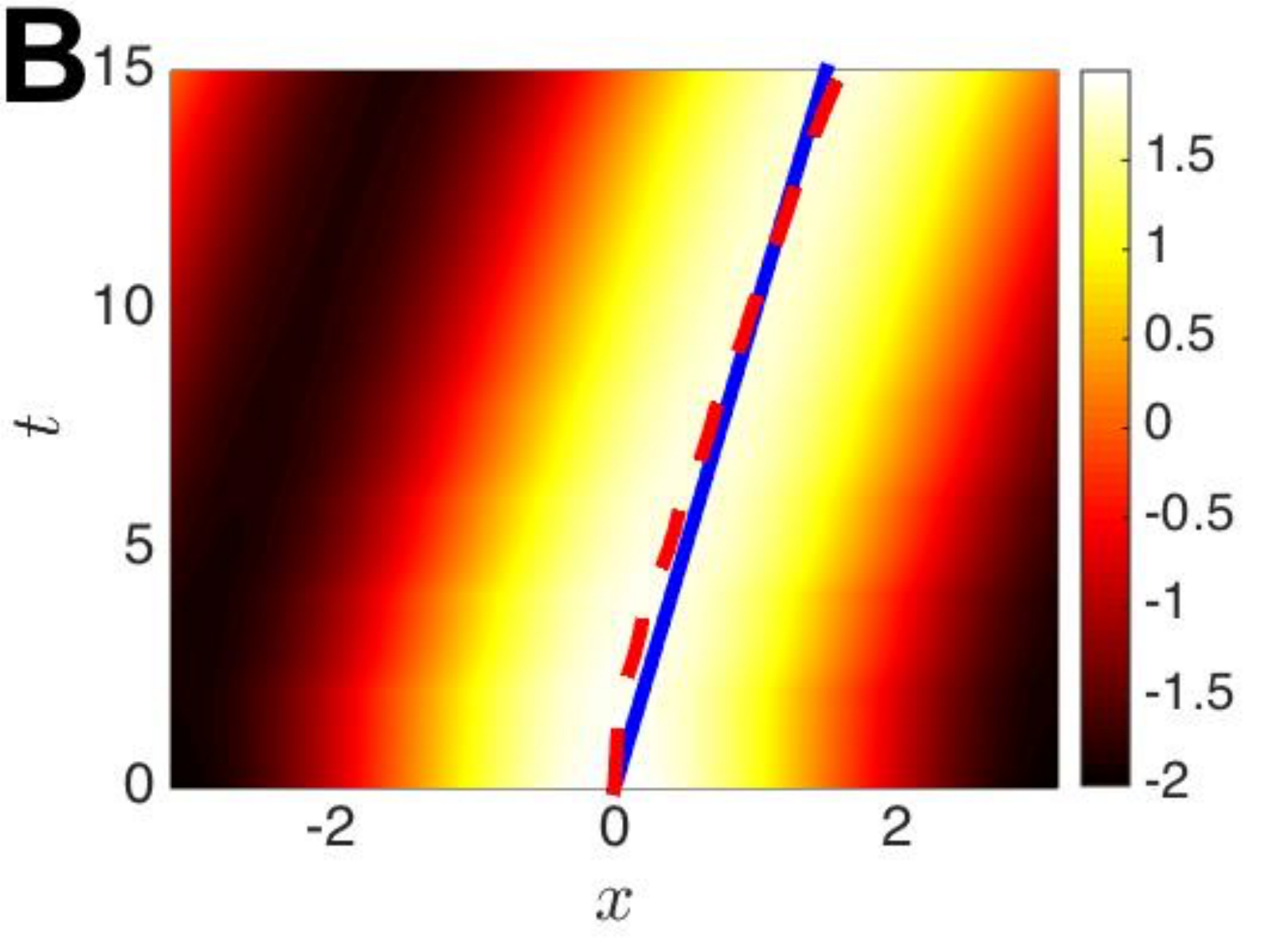}
\includegraphics[width=5cm]{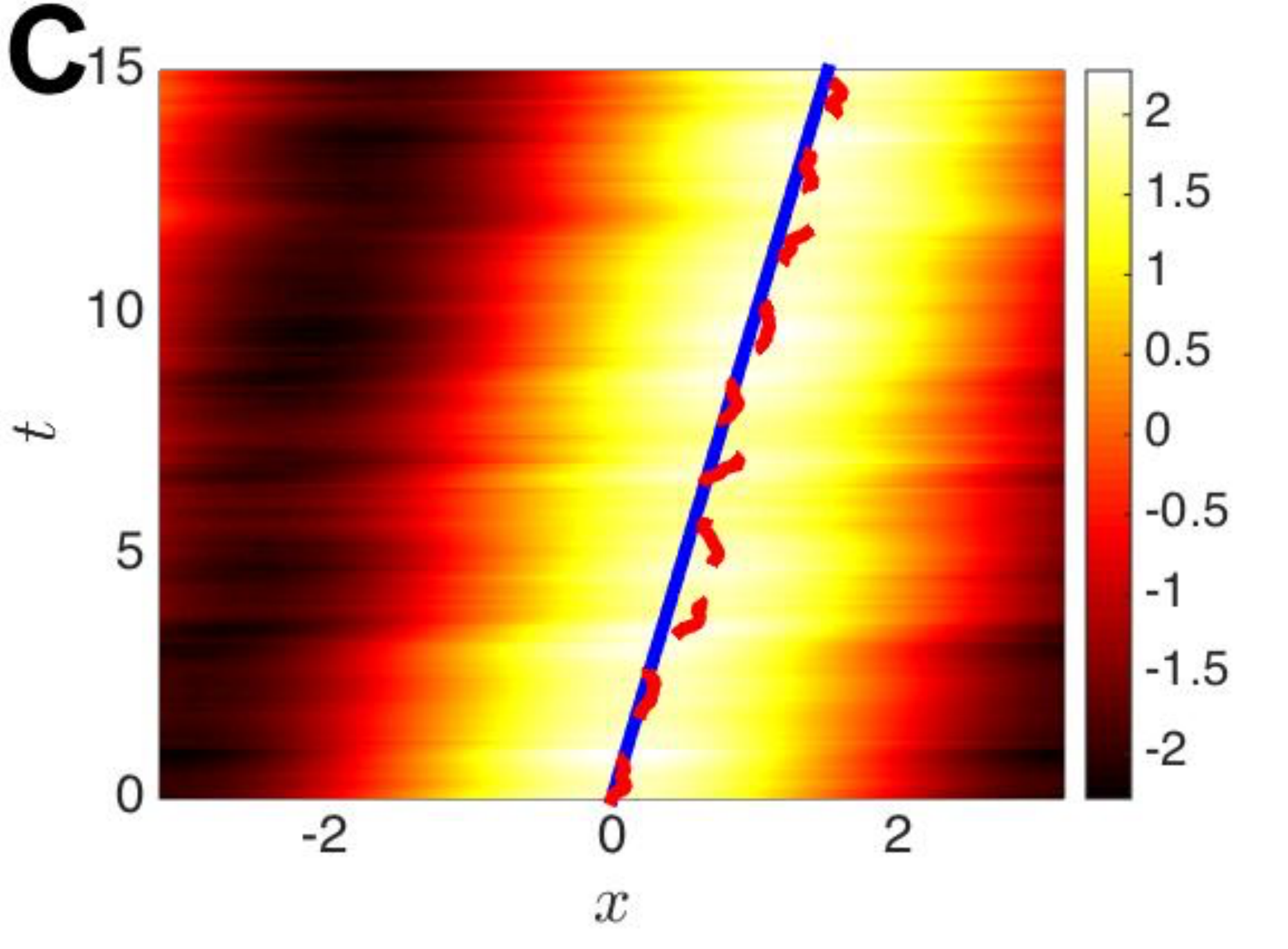}
\end{center}
\caption{Discrete control reduces error in imperfect velocity-integrating networks. ({\bf A}) Activity plot $u(x,t)$ shows perfect velocity-integrating network is unaffected by discrete control given by (\ref{disccontr}), since it accumulates no errors. ({\bf B}) Spatially heterogeneous network with $\sigma = 0.1$ and $w_u(x) = \sin (x)$ is corrected by discrete control so that the represented position (dashed line) is much closer to the true position (solid line) than in the uncorrected case (compare with Fig. \ref{fig2}{\bf B}). ({\bf C}) Network perturbed by noise ($\epsilon = 0.2$ and $C(x) = \cos (x)$) has its represented position corrected by discrete control (compare with Fig. \ref{fig2}{\bf C}). Discrete control (\ref{discexp}) is given at intervals $t_{k+1} -t_k = \Delta t = 2$ with strength $\lambda = 1$ and decay timescale $\tau = 1$. Other parameters and numerical simulations are as in Fig. \ref{fig2}.}
\label{fig3}
\end{figure}
 
We demonstrate the impact of discrete control (\ref{disccontr}) on the dynamics of neural fields that imperfectly integrate their velocity inputs, comparing to a perfectly integrating network for reference (Fig. \ref{fig3}{\bf A}). Integrating (\ref{disccontr}), we find that the discrete perturbations to the velocity signal are given by a series of exponentially decaying impulses
 \begin{align}
 v_c(t) = \lambda \sum_{k=1}^{N_c} r(t_k) \e^{-(t-t_k)/\tau} H(t- t_k). \label{discexp}
 \end{align}
Furthermore, note in the limit of the timestep $t_{k+1}-t_k$ between subsequent cues tending to zero and $\tau \to 0$, the equation for discrete control (\ref{disccontr}) approaches that of the continuous control (\ref{contcontr}). Assuming for demonstration that cues are spaced in such a way that an animal encounters one every 2 units of time ($t_{k+1}-t_k = 2$ for all $k=1,...,N_c-1$), we demonstrate in a single realization that a network with spatially heterogeneous coupling can recover its ability to correctly integrate velocity (Fig. \ref{fig3}{\bf B}). In a similar way, networks with additive noise can have their velocity integration corrected by the sensory feedback signal given by discrete control impulses (\ref{discexp}) as shown in Fig. \ref{fig3}{\bf C}. In the next section, we will analyze the impact of heterogeneity and noise on the position of the bump in a low-dimensional approximation of the bump's center of mass $\Delta (t)$.

\section{Analysis and low-dimensional reduction of bump solutions}
\label{potwell}

To develop an understanding of the impact the sensory feedback signal has on the statistics of bump position in (\ref{nfmodel}), we derive a low-dimensional approximation of the model that projects the dynamics down to a single equation describing bump position $\Delta (t)$. Our analysis is adapted from recent studies of stochastic neural field equations assuming the impact of perturbations to a translationally symmetric neural field can be separated into slow timescale changes to the position of patterns along with fast timescale changes to the profile of patterns \citep{bressloff12b,kilpatrick13,bressloff15}. Such analysis must begin by constructing the patterns of the unperturbed system. In our case we take the velocity inputs, heterogeneity, asymmetry, and noise all to be perturbations to a translationally symmetric system (taking $v = \sigma = \phi = \epsilon = 0$ in (\ref{nfmodel})). While it is possible to develop exact analytical results in the case wherein we break the symmetry of this model, which we show, it is also helpful to collect the effects of all the possible perturbations to (\ref{nfmodel}) into a single scalar equation. Doing so makes it more clear how such perturbations interact.

\subsection{Stationary bump solutions to the translation symmetric network}

We begin by assuming the homogeneous connectivity function $w_0(x)$ in (\ref{wf}) satisfies evenness ($w_0(x) = w_0(-x)$) and there is no heterogeneity ($\sigma = 0$) or asymmetry ($\phi = 0$) in $w(x,y)$. In this case, it is possible to show there is a stationary bump solution $u(x,t) = U(x)$ with $U(x)> \theta$ over an excited region $x \in [a_1,a_2]$ \citep{amari77,ermentrout98} in the absence of velocity inputs ($\tilde{v} \equiv 0$) in (\ref{nfmodel}). Furthermore, the weight function is translation symmetric since
\begin{align}
w_0((x+s)-(y+s)) = w_0(x-y + s-s) = w_0(x-y),
\end{align}
so there will be a continuum of bump locations associated with any single bump solution to (\ref{nfmodel}) in this case. Stationary bumps satisfy the equation
\begin{align}
U(x) = \int_{- \pi}^{\pi} w_0(x-y) f(U(y)) dy. \label{beqn}
\end{align}
Note that $U(x + s)$ will also be a solution for any $s$, since
\begin{align*}
U(x+s) = \int_{- \pi}^{\pi} w_0(x-y) f(U(y+s)) dy.
\end{align*}
A change of variables $y+s \mapsto z$ yields
\begin{align*}
U(x+s) = \int_{- \pi}^{\pi} w_0(x+s - z) f(U(z)) dz,
\end{align*}
and another change of variables $x + s \mapsto x'$ yields
\begin{align*}
U(x') = \int_{- \pi}^{\pi} w_0(x' -z ) f(U(z)) dz,
\end{align*}
which is precisely the equation (\ref{beqn}). Now, taking the high gain limit $\gamma \to \infty$, we employ the Heaviside firing rate function (\ref{H}). Doing so allows us to generate an equation for the bump width $d = a_2 - a_1$ as in \cite{amari77}. In this case, the equation (\ref{beqn}) becomes
\begin{align*}
U(x) = \int_{a_1}^{a_2} w_0(x-y) dy.
\end{align*}
We then use the threshold crossing conditions $U(a_1) = U(a_2) = \theta$ and evenness of $w_0(x)$ to derive
\begin{align*}
U(a_1) = \int_{a_1}^{a_2} w_0(a_1 - y) dy = \int_0^{a_2 - a_1} w_0(-z) dz = \int_0^d w_0(z) dz =& \theta \\
U(a_2) = \int_{a_1}^{a_2} w_0(a_2 - y) dy = - \int_{a_2-a_1}^0 w_0(z) dz = \int_0^d w_0(z) dz =& \theta. 
\end{align*}
Note that the evenness of $w_0(x)$ allows us to manipulate the above equations so they are the same equalities. If evenness did not hold, the above pair of equations would each be different and we would have an overdetermined system for the bumpwidth $d$, meaning stationary bumps do not exist. Thus,
\begin{align*}
W(d) = \int_0^d w_0(x) dx = \theta \hspace{4mm} \Rightarrow \hspace{4mm} d = W^{-1} ( \theta).
\end{align*}
For example, in the case of a cosine weight $w_0(x) = \cos (x)$, we have
\begin{align}
W(d) = \int_0^d \cos (x) dx = \sin (d) = \theta \hspace{4mm} \Rightarrow \hspace{4mm} d = \sin^{-1} \theta, \pi - \sin^{-1} \theta.  \label{bwidcos}
\end{align}
As mentioned, the threshold conditions specify the width $d$ of the bump. Translation symmetry allows the position of the bump to be anywhere $x \in [- \pi, \pi]$, which allows this network to integrate and store velocity inputs as a position memory. As mentioned in section \ref{model}, the position of the bump will be given by its center of mass (\ref{combump}), which for unimodal and even symmetric bumps will also be given by the peak
\begin{align}
\Delta = {\rm argmax}_x U(x). \label{amax}
\end{align}
For example, in the case of cosine weight functions $w_0(x) = \cos (x)$, there is an even symmetric solution such that $a_1 = -a$ and $a_2 = a$, so
\begin{align}
U(x) = \int_{-a}^a \cos (x-y) dy = 2 \sin (a) \cos (x).  \label{bsol}
\end{align}
Thus, the location of the bump as computed by (\ref{amax}) is $\Delta = 0$. Similarly, if we compute the center of mass using (\ref{combump}), we find
\begin{align*}
\Delta = \int_{- \pi}^{\pi} x f(U(x)) dx = \int_{- a}^{a} x dx = 0,
\end{align*}
which is consistent.

\subsection{Perfect velocity integration by traveling bumps}

Now we explore the impact of velocity inputs ($v(t) \neq 0$) on the translationally symmetric network ($w(x,y) = w_0(x-y)$). For now, we assume constant velocity inputs, $v(t) \equiv v_0$. Assuming the bump subsequently moves at a constant speed $c$, we look for a traveling wave solution $u(x,t) = U(\xi)$ where $\xi = x- ct$. We will show that the traveling wave speed $c$ is exactly equal to the velocity input amplitude $v_0$, under the assumption that $w_v(x) = -w_0'(x)$ in (\ref{nfmodel}). Plugging these conditions into (\ref{nfmodel}), we find
\begin{align*}
-c U'(\xi) + U(\xi) = \int_{- \pi}^{\pi} \left[ w_0(\xi - y) + v_0 w_v(\xi - y) \right] f(U(y)) dy.
\end{align*}
Now plugging in our requirement that the velocity portion of the weight function $w_v(x) = -w_0'(x)$, we have
\begin{align}
-c U'(\xi) + U(\xi) = - v_0 \int_{- \pi}^{\pi} w_0'(\xi - y) f(U(y)) dy + \int_{- \pi}^{\pi} w_0(\xi - y) f(U(y)) dy. \label{waveeqn}
\end{align}
Under the assumption that the function $U(\xi)$ satisfies the equality (\ref{beqn}), we have differentiate this equation to yield
\begin{align}
U'(\xi) = \int_{- \pi}^{\pi} w_0'(\xi - y) f(U(y)) dy. \label{Upeqn}
\end{align}
Canceling the (\ref{beqn}) portion of (\ref{waveeqn}), we find that
\begin{align}
c U'(\xi) = v_0 \int_{- \pi}^{\pi} w_0'(\xi - y) f(U(y)) dy. \label{cv0eqn}
\end{align}
The equality (\ref{cv0eqn}) follows from (\ref{Upeqn}) as long as we set $c \equiv v_0$. Another implication of our analysis is that the shape of the bump $U(\xi)$ will be the same no matter what $c$ (equivalently $v_0$) is, suggesting there will be no relaxation time if the external drive $v_0$ were to be changed abruptly. In this way, we can expect the translation symmetric version of the network (\ref{nfmodel}) to integrate inputs perfectly as was originally proposed by \cite{zhang96}.

\subsection{Imperfect integration due to heterogeneity, asymmetry, and noise}

\begin{figure}
\begin{center} \includegraphics[width=7.2cm]{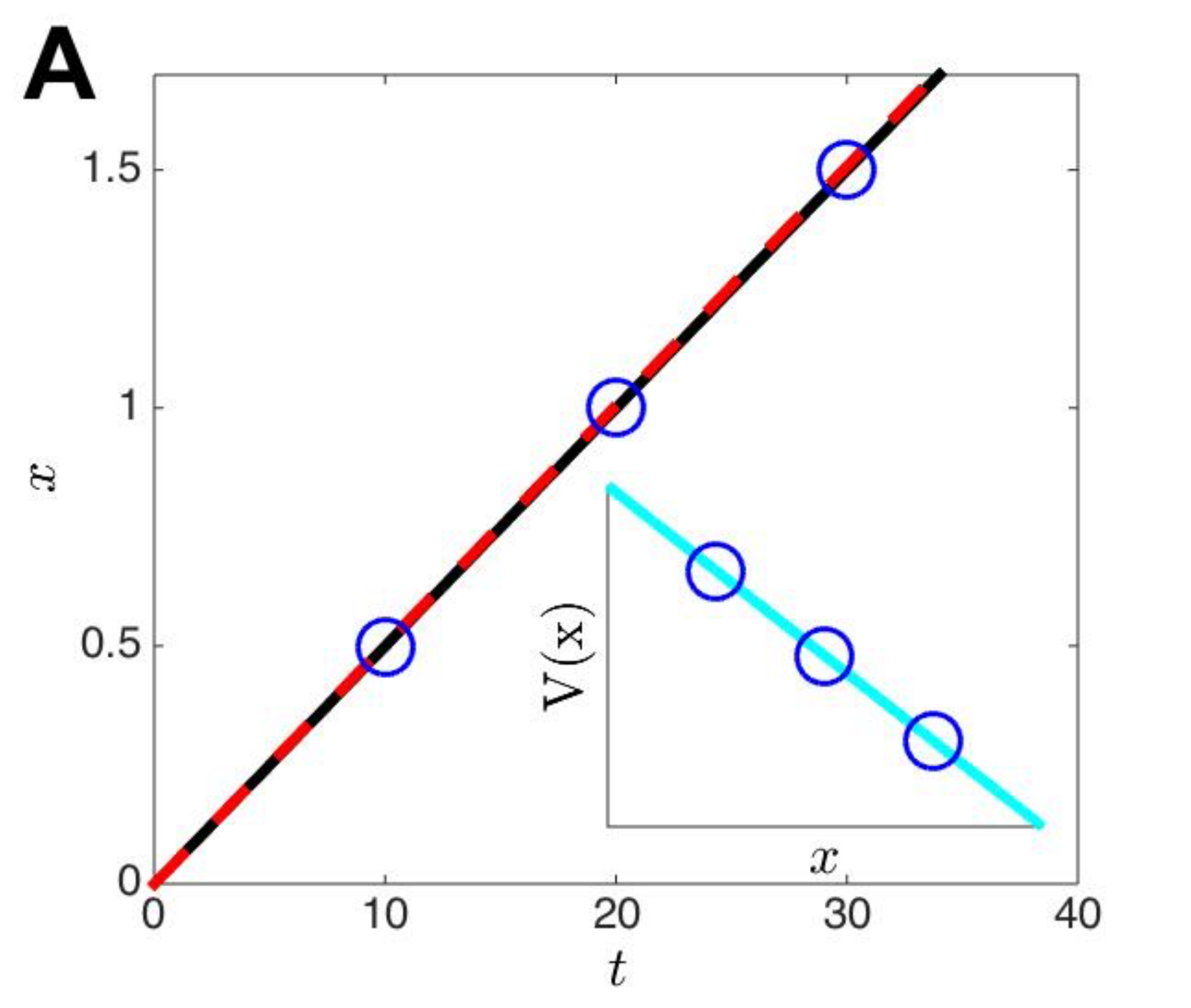}
\includegraphics[width=8cm]{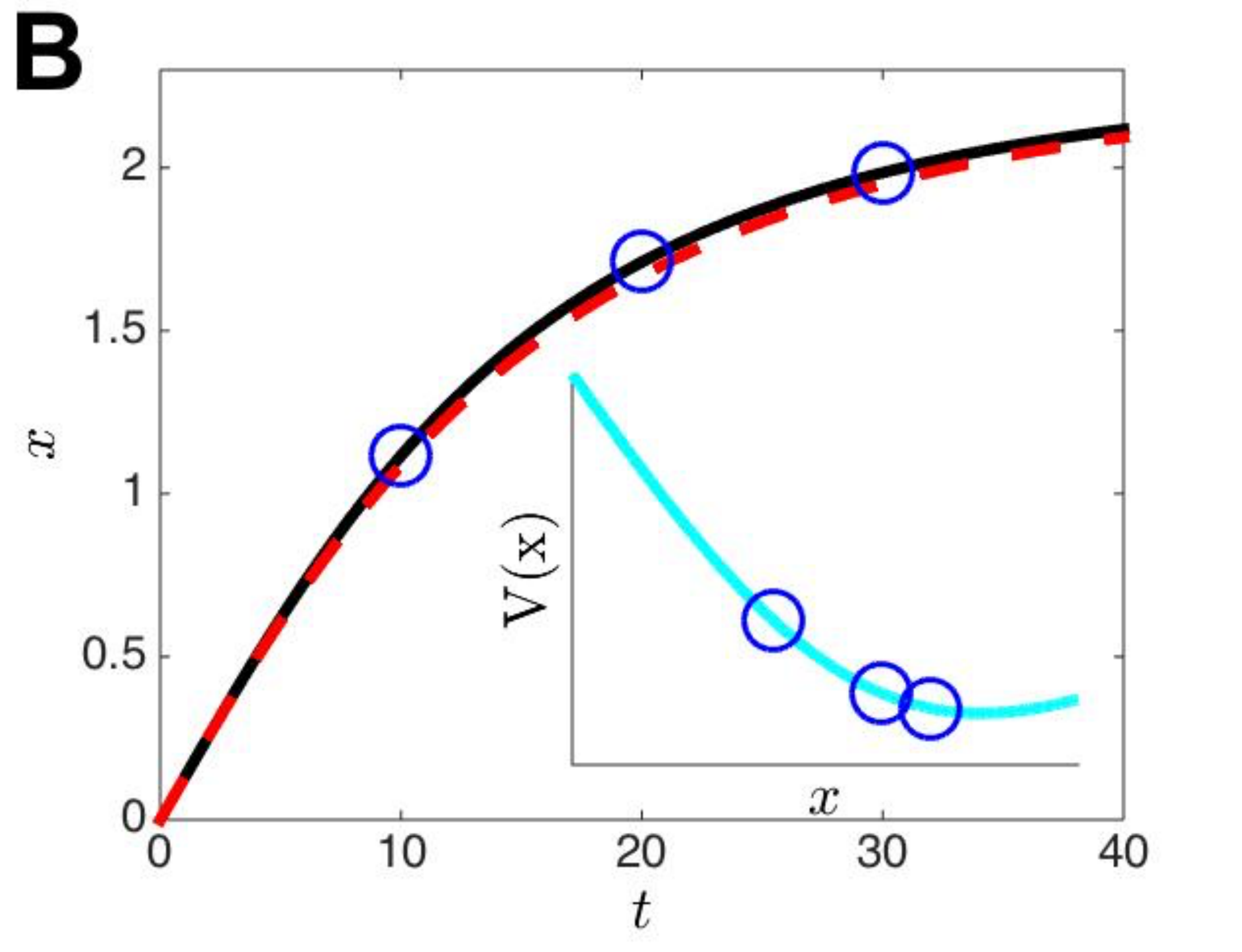} \\ \includegraphics[width=7cm]{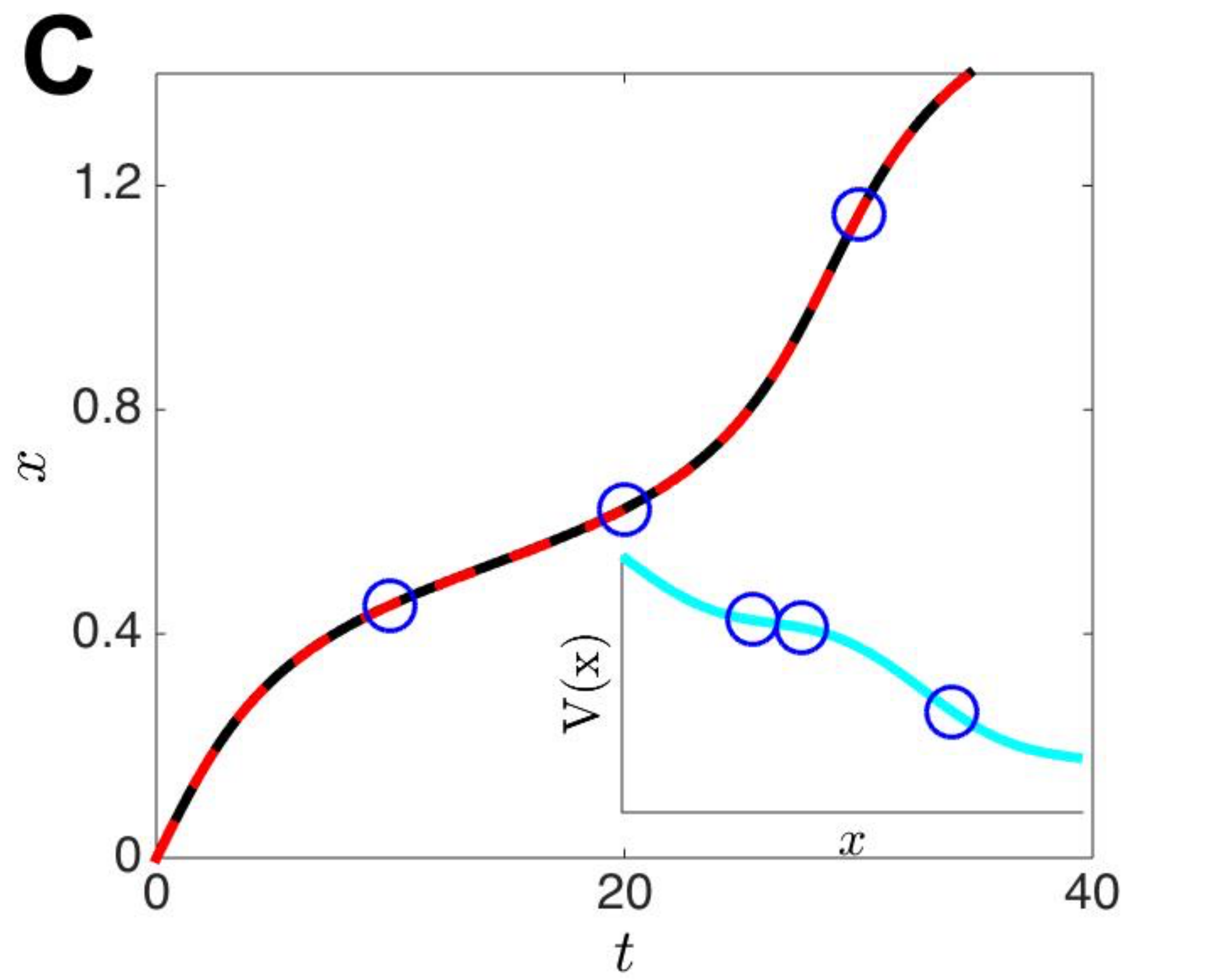}
\includegraphics[width=8cm]{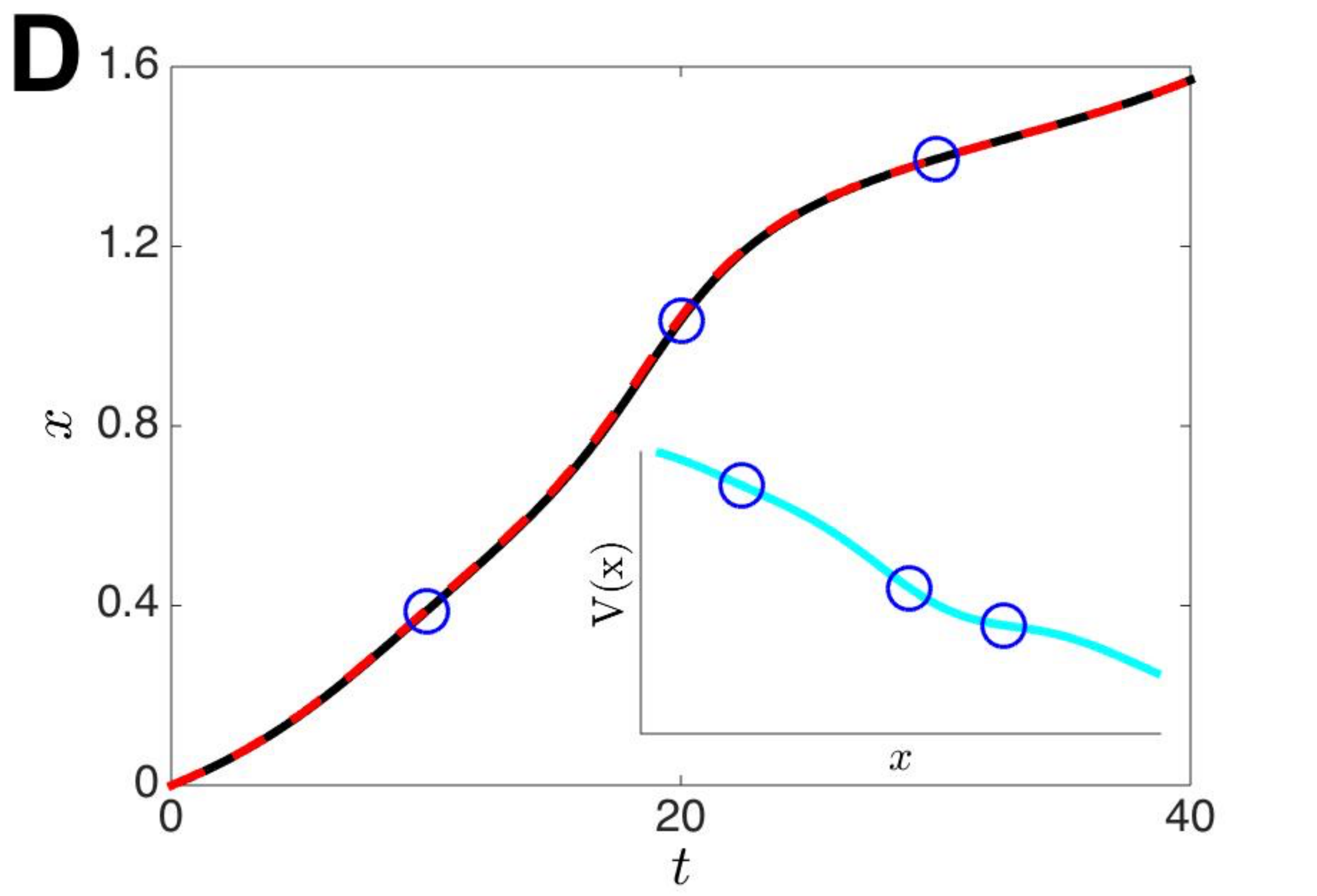} \end{center}
\caption{The low-dimensional equation (\ref{ddel2}) for the bump position $\Delta (t)$ provides an accurate approximation of the bump's movement in the full neural field model (\ref{nfmodel}). ({\bf A}) Perfect integration of the constant velocity input $v(t) = 0.05$ leads to a constantly drifting bump (solid line) whose position is well approximated by the projected variable $\Delta (t)$ (dashed line). Inset shows the tilted potential $V(\Delta)$ resulting from the constant velocity input. Circles provide corresponding locations between the two plots at $t=10,20,30$. ({\bf B}) Spatial heterogeneity $w_u(x) = \sin (x)$ with $\sigma = 0.1$ causes bumps to drift toward local attractors of the network. Inset shows potential with a local minimum to which the trajectory is attracted. ({\bf C}) Spatial heterogeneity $w_u(x) = \sin (6x)$ with $\sigma = 0.2$ leads to a more rapid oscillation in the trajectory $\Delta (t)$. ({\bf D}) Spatial heterogeneity $w_u(x) = \sin (4x) + \cos (8x)$ with $\sigma = 0.1$ leads to a less regular deviation in the trajectory $\Delta (t)$. Heaviside firing rate function (\ref{H}) has threshold $\theta = 0$. Numerical simulations are run using the same parameters as in Fig. \ref{fig2}.}
\label{fig4}
\end{figure}

Now that we have explored the dynamics of the perfect velocity-integrating network, we study the impact of introducing heterogeneities ($\sigma$), asymmetry ($\phi$), and noise ($\epsilon$) into the network (\ref{nfmodel}). Rather than deriving exact solutions as we did for the translationally symmetric system, we take a perturbative approach under the assumption that alterations to the symmetric system are weak. Following perturbation methods originally developed for the study of front propagation in reaction-diffusion systems \citep{panja04,sagues07}, we employ a separation of time scales to decompose these effects into a slowly evolving displacement $\Delta (t)$ of the bump from its uniformly translating position and perturbations to the bump profile $\Phi (x,t)$. This yields the following decomposition
\begin{align}
u(x,t) = U (x - \Delta (t)) + \epsilon \Phi (x  - \Delta (t), t) + {\mc O}(\epsilon^2), \label{pertans}
\end{align}
where we assume $\sigma, \phi, \tilde{v}(t) \sim {\mc O}(\epsilon)$. Plugging the ansatz (\ref{pertans}) into (\ref{nfmodel}) and expanding in powers of $\epsilon$, we find that at ${\mc O}(1)$, the system has a stationary bump solution given by (\ref{beqn}). At linear order ${\mc O}(\epsilon)$, we find the following equation
\begin{align}
\epsilon d \Phi (x,t) =&  \epsilon {\mc L} \Phi (x,t) dt + U'(x) \d \Delta (t)  + \sigma \int_{- \pi}^{\pi} w_u(y+\Delta(t)) w_0(x-y) f(U(y)) dy dt \label{oepsphi} \\ &  - (\tilde{v}(t) + \phi) \int_{- \pi}^{\pi} w_0'(x-y) f(U(y)) dy dt  + \epsilon d W(x,t) \nonumber
\end{align}
where $\mathcal{L}$ is a linear functional given by 
\begin{equation*}
\mathcal{L} p(x)  := - p(x) + \int_{-\pi}^\pi w_0(x-y) f'(U(y)) p(y) dy
\end{equation*}
and its adjoint operator
\begin{equation*}
\mathcal{L^*} q(x)  = - q(x) + f'(U(x)) \int_{-\pi}^{\pi} w_0(x-y) q(y) dy.
\end{equation*}
To ensure a solution to (\ref{oepsphi}), we require that the inhomogeneous portion of the equation be orthogonal to the nullspace of the adjoint operator ${\mc L}^*$. Indeed, the nullspace of ${\mc L}^*$ is spanned by $\varphi(x) = f'(U(x)) U'(x)$, where $U(x)$ is defined by (\ref{beqn}), which we can verify using integration of parts
\begin{equation}  \label{nulleqn}
\begin{aligned}
\mathcal{L^*} \varphi(x) &= -\varphi(x) + f'(U(x)) \int_{-\pi}^{\pi} w_0(x-y) \varphi (y) dy \\
&= f'(U(x)) \Big(  -U'(x) +  \int_{-\pi}^{\pi} w_0(x-y) f'(U(y)) U'(y) dy \Big) \\
&= f'(U(x)) \Big( -U'(x) + \int_{-\pi}^{\pi} \frac{d}{dy} \big( w_0 (x-y) ) f(U(y)) dy \Big) \\
&= f'(U(x)) \Big( -U'(x) + \frac{d}{dx} \Big( \int_{-\pi}^{\pi}  w_0 (x-y) f(U(y)) dy \Big) \Big) = 0.
\end{aligned}
\end{equation}
The last line holds by differentiating the bump existence equation as in (\ref{Upeqn}). Now, by taking inner products of the null vector $\varphi$ with the ${\mc O}(\epsilon)$ equation (\ref{oepsphi}), we can derive an evolution equation for $\Delta(t)$, the position of the bump
\begin{align}
- \int_{- \pi}^{\pi} \varphi (x) U'(x) dx d \Delta (t) = & \sigma \int_{- \pi}^{\pi} \varphi (x) \int_{- \pi}^{\pi} w_u(y+ \Delta (t)) w_0(x-y) f(U(y)) dy dx dt  \label{ddel1} \\
& - (\tilde{v}(t) + \phi) \int_{- \pi}^{\pi} \varphi (x) U'(x) dx dt + \epsilon \int_{ - \pi}^{\pi} \varphi (x) d W(x,t) dx,  \nonumber
\end{align}
where we have applied the equation (\ref{Upeqn}). We can simplify the equation (\ref{ddel1}) further by isolating $d \Delta (t)$ to yield the stochastic differential equation
\begin{align}
d \Delta (t) = \left[ F(\Delta (t)) + v(t) + v_c(t) + \phi \right] dt + d {\mc W}(t), \label{ddel2}
\end{align}
where the impact of synaptic spatial heterogeneities is described by the nonlinear function
\begin{align}
F(\Delta ) =  - \sigma \frac{ \int_{- \pi}^{\pi} f'(U(x)) U'(x) \int_{- \pi}^{\pi} w_u(y + \Delta ) w_0(x-y) f(U(y)) dy dx}{\int_{- \pi}^{\pi} f'(U(x)) U'(x)^2 dx},  \label{schetero}
\end{align}
and the noise term has been projected to a temporal white noise process ${\mc W}(t)$ with mean zero ($\langle {\mc W}(t) \rangle = 0$) and variance $\langle {\mc W}(t)^2 \rangle = Dt$ with associated diffusion coefficient
\begin{align}
D = \epsilon^2 \frac{\int_{- \pi}^{\pi} \int_{- \pi}^{\pi} f'(U(x))U'(x) f'(U(y)) U'(y) C(x-y) dy dx}{\left[ \int_{- \pi}^{\pi} f'(U(x)) U'(x)^2 dx \right]^2}.  \label{dcoef}
\end{align}
Setting $v_c(t) \equiv 0$ and $v(t) \equiv v_0$ (constant), the dynamics of the position variable $\Delta (t)$ can be equivalently described by a potential function
\begin{align}
V( \Delta ) = - \int \left[ F(\Delta) + v_0 + \phi \right] d \Delta = - \int F( \Delta ) d \Delta - (v_0 + \phi) \Delta, \label{potent}
\end{align}
so $\Delta (t)$ will descend the gradient of $V(\Delta)$ toward its local minima. Note that in the case $F(\Delta) \equiv \phi \equiv {\mc W}(t) \equiv 0$, the control term will vanish $v_c(t) \equiv 0$ and the bump will perfectly integrate the velocity input, $\Delta (t) = \int_0^t v(s) ds$. We find that the low dimensional approximation is in excellent agreement with simulations of the full system in this case of perfect integration (Fig. \ref{fig4}{\bf A}).

Ignoring the control for the time being in (\ref{ddel2}), we can also identify how different network imperfections contribute to the resultant error in path integration. To do so, we simply compute the error function $r(t)= \Delta_T(t) - \Delta (t)$ as given in (\ref{err}). First, note that in a network with asymmetry $\phi \neq 0$ and no heterogeneity $F(\Delta) \equiv 0$, the long term error accumulates linearly in time
\begin{align*}
r(t) = \Delta_T (t) - \Delta (t) = \int_0^t v(s) ds - \int_0^t \left[ v(s) + \phi \right] ds = - \phi t,
\end{align*}
so the animal's true position $\Delta_T$ will be behind (in front of) the estimated position $\Delta$ when $\phi >0$ ($\phi < 0$). We will demonstrate the impact external control via sensory cues has upon this error in section \ref{control}. Errors due to arbitrary heterogeneities are more difficult to express explicitly. In general, we can express the solution to (\ref{ddel2}) implicitly in this case if we assume the velocity is constant $v(t) \equiv v_0$:
\begin{align}
G(\Delta (t)) = \int_0^{\Delta} \frac{d y}{F(y) + v_0} = t.
\end{align}
If indeed the function $G(\Delta)$ is invertible, then we can express $\Delta (t) = G^{-1} (t)$, so
\begin{align}
r(t) = \Delta_T(t) - \Delta (t) = v_0 t - G^{-1} (t).
\end{align}
We will demonstrate some cases where we can perform this calculation explicitly in subsection \ref{explicit}. Note that the main impact of heterogeneities is to establish a finite number of discrete attractors, in the velocity input-free system, so that bumps drift toward these locations (Fig. \ref{fig4}{\bf B}). Even in the velocity-driven network, spatial heterogeneities lead to a sinuous trajectory of the bump that is mismatched to a straight integration of velocity input (Fig. \ref{fig4}{\bf C},{\bf D}). Lastly, note that the impact of noise can be quantified by averaging across realizations of the stochastic process
\begin{align*}
\Delta (t) = \int_0^t v(s) ds + {\mc W}(t) = \Delta_T(t) + {\mc W}(t).
\end{align*}
While the mean position will be the same for the true and encoded positions ($\Delta_T(t) - \langle \Delta (t) \rangle = 0$), the variance will grow linearly in time
\begin{align*}
\langle r(t)^2 \rangle = \langle (\Delta_T(t) - \Delta(t))^2 \rangle = \langle {\mc W}(t)^2 \rangle = Dt,
\end{align*}
where the diffusion coefficient $D$ can be computed using (\ref{dcoef}). Previous work has characterized the impact of the bump profile and spatiotemporal noise correlation structure on the diffusion coefficient $D$, providing some explicit calculations \citep{kilpatrick13}. In general, the main effects of noise perturbations on the bump will be experienced by the bump edges, where the activity variable $u(x,t)$ crosses the firing rate threshold $\theta$. We now provide some explicit calculations demonstrating the impact of spatial heterogeneity on the long term position of the bump.

\subsection{Explicit results for spatially heterogeneous networks with a Heaviside firing rate}
\label{explicit}

Several previous studies have characterized the impact of periodic microstructure on the propagation of waves in neural media \citep{bressloff01,kilpatrick08,coombes11}. Typically, periodic heterogeneities can slow down waves and even cause propagation failure. We extend these previous results here, showing that the low dimensional equation (\ref{ddel2}) allows us to estimate the location of bifurcations separating detectable and undetectable constant velocity inputs $v(t) \equiv v_0$. Again, we are ignoring the impact of control at this point, studying its effects in more detail in section \ref{control}. To allow for fully general weight heterogeneities, we consider the decomposition given by (\ref{ewf}). Thus, we can integrate each of the Fourier modes independently to derive the function $F(\Delta)$ given by (\ref{schetero}). Furthermore, we assume a cosine for the homogeneous weight function $w_0(x) = \cos (x)$ and a Heaviside firing rate (\ref{H}).

To begin, note that the bump solution is given by (\ref{bsol}) and the half-width is $a = \frac{1}{2} \left[ \pi - \sin^{-1} \theta  \right]$ as given by (\ref{bwidcos}). Therefore, the spatial derivative $U'(x) = - 2 \sin (a) \sin (x)$. Furthermore, the null vector defined by (\ref{nulleqn}) is spanned by the difference of delta distributions $\delta (x+a) - \delta (x-a)$. This means that the frequency $n$ cosine Fourier components of the heterogeneity, with scaling $\alpha_n$, contribute the function $F(\Delta)$ in the following way
\begin{align}
F_{\alpha_n} (\Delta) =& \frac{\int_{- \pi}^{\pi} \left( \delta (x+a) - \delta (x-a) \right) \int_{-a}^{a} \cos (n(y + \Delta)) \cos (x-y) dy dx}{ 2 \sin a \int_{- \pi}^{\pi} \left( \delta (x+a) - \delta (x-a) \right) \sin x dx  } \nonumber \\
=& -\frac{\int_{- \pi}^{\pi} \left( \delta (x+a) - \delta (x-a) \right) (n \cos (a) \sin (na) - \sin (a) \cos (na)) \cos (x) dx \cos (n \Delta)}{ 2 (n^2-1) \sin^2 a } \nonumber \\
& + \frac{\int_{- \pi}^{\pi} \left( \delta (x+a) - \delta (x-a) \right) ( \cos (a) \sin (na) - n \sin (a) \cos (na)) \sin (x) dx \sin (n \Delta)}{ 2 (n^2-1) \sin^2 a } \nonumber \\
=& \frac{n \cos (na) -  \cot (a) \sin (na)}{n^2-1} \sin (n \Delta). \nonumber
\end{align}
In a similar way, we can compute the coefficients arising from the sine Fourier components with scaling $\beta_n$ as
\begin{align}
F_{\beta_n}(\Delta) =& \frac{\int_{- \pi}^{\pi} \left( \delta (x+a) - \delta (x-a) \right) \int_{-a}^{a} \sin (n(y + \Delta)) \cos (x-y) dy dx}{ 2 \sin a \int_{- \pi}^{\pi} \left( \delta (x+a) - \delta (x-a) \right) \sin x dx  }  \nonumber \\
=&  -\frac{\int_{- \pi}^{\pi} \left( \delta (x+a) - \delta (x-a) \right) (n \cos (a) \sin (na) - \sin (a) \cos (na)) \cos (x) dx \sin (n \Delta)}{ 2 (n^2-1) \sin^2 a }  \nonumber \\ & - \frac{\int_{- \pi}^{\pi} \left( \delta (x+a) - \delta (x-a) \right)( \cos (a) \sin (na) - n \sin (a) \cos (na)) \sin (x) dx \cos (n \Delta)}{ 2 (n^2-1) \sin^2 a } \nonumber \\
= & \frac{\cot (a) \sin (na) - n \cos (na)}{n^2 -1} \cos (n \Delta). \nonumber
\end{align}
Thus, we can write the resultant heterogeneity in general as
\begin{align}
F( \Delta)  = \sigma \sum_{n=1}^N {\mc C}_n \left[ \alpha_n \sin (n \Delta) - \beta_n \cos (n \Delta) \right],  \label{hetfour}
\end{align}
where
\begin{align}
{\mc C}_n = \frac{n \cos (na) -  \cot (a) \sin (na) }{n^2-1},
\end{align}
and notice in the special case $n=1$, we can take the limit $n \to 1$ to find
\begin{align}
{\mc C}_1 = \frac{\sin (a) \cos (a) - a}{2 \sin (a)}.
\end{align}

\begin{figure}
\begin{center} \includegraphics[width=7.5cm]{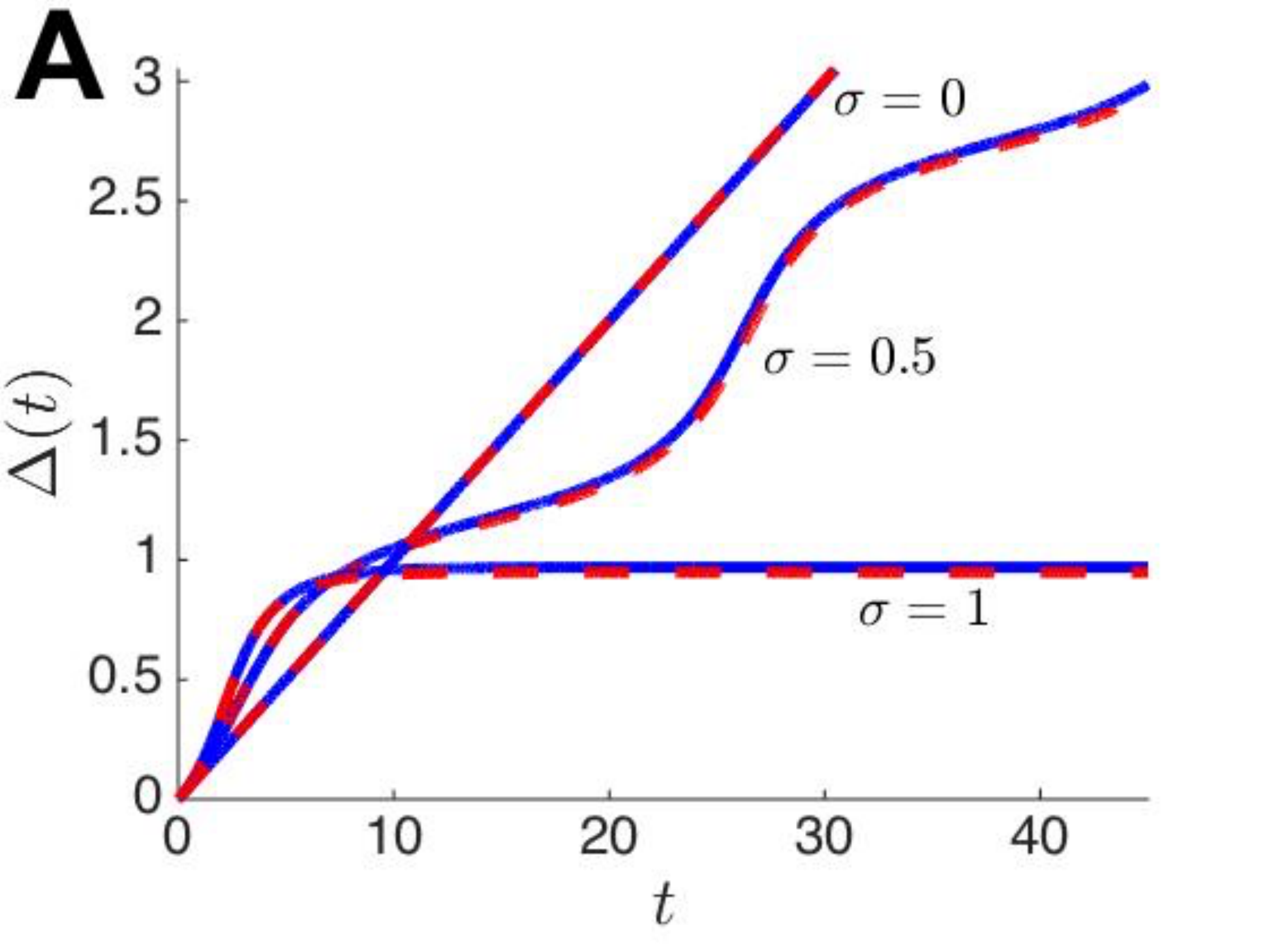} \includegraphics[width=7.5cm]{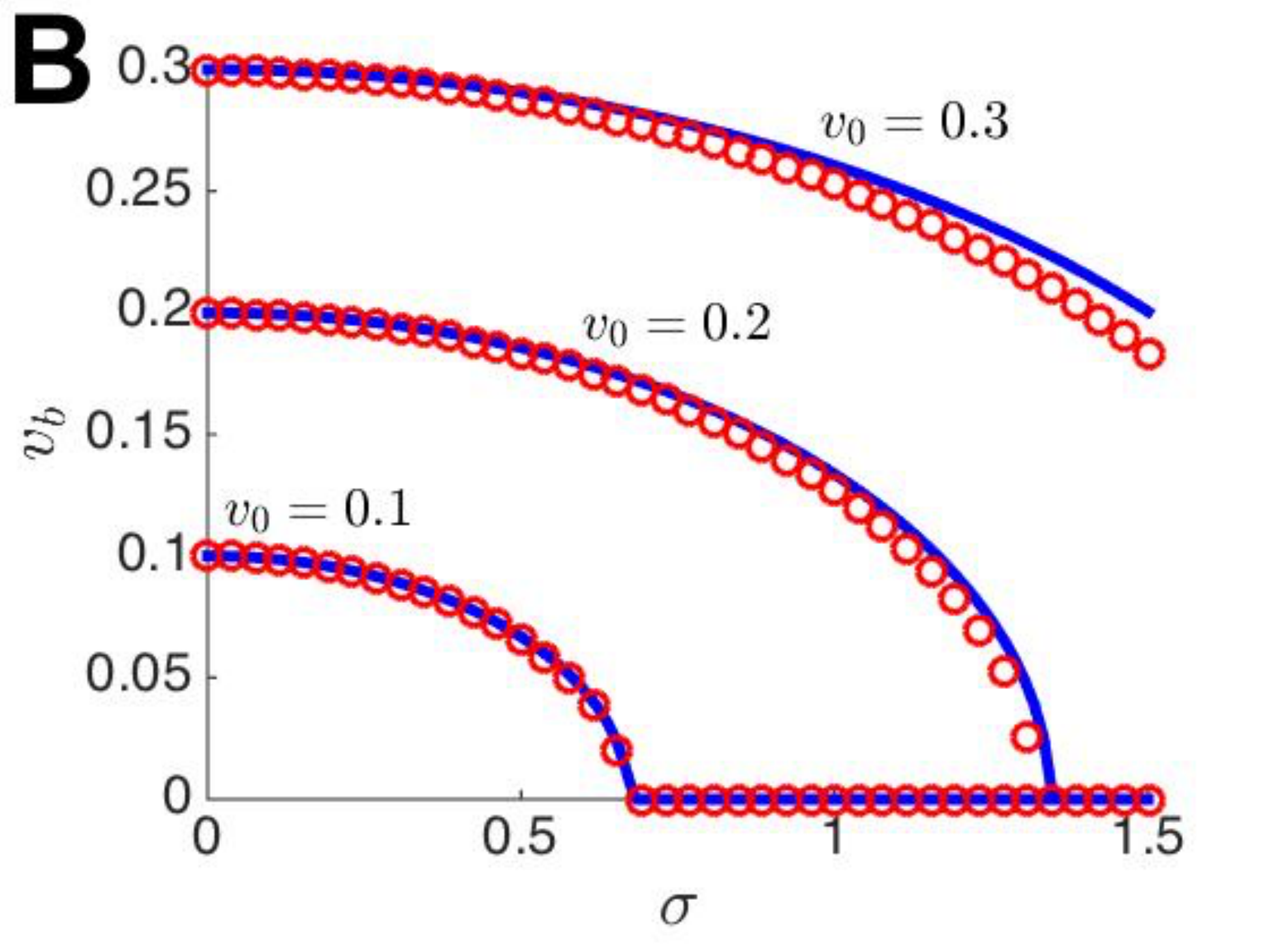} \end{center}
\caption{Spatial heterogeneity slows and even stops the propagation of velocity-driven bumps. ({\bf A}) Bump position $\Delta (t)$ demonstrates the variant propagation velocity occurring when heterogeneity ($\sigma = 0.5$) is introduced as opposed to the homogeneous network ($\sigma = 0$). For strong enough heterogeneity ($\sigma = 1$), propagation fails. Theory (solid line) given by (\ref{deltan}) is well matched to simulations (dashed line) of the full model (\ref{nfmodel}). ({\bf B}) Bump velocity $v_b$ decreases as a function of heterogeneity strength $\sigma$ until propagation failure occurs at $\sigma = (m^2 -1) |v_0|/|\cot (a) \sin (ma) - m \cos (ma)|$. Firing rate function is Heaviside (\ref{H}), heterogeneity is $w_u(x) = \cos (mx)$, and other parameters $\theta = 0.5$ and $m=4$.}
\label{fig5}
\end{figure}

We can explicitly compute the solution to (\ref{ddel2}) in some special cases of the heterogeneity $F(\Delta)$, defined by (\ref{hetfour}). In particular, we focus on a single cosine-shaped heterogeneity so that $\alpha_m=1$, $\alpha_n = 0$ ($n\neq m$), and $\beta_n = 0$ ($\forall n$). Furthermore, we assume a constant input velocity $v(t) \equiv v_0$, so that the scalar equation (\ref{ddel2}) for $\Delta (t)$ is given by
\begin{align}
\dot{\Delta}(t) = \kappa \sin (m \Delta ) + v_0, \label{ddelsine}
\end{align}
where $\kappa = \sigma {\mc C}_n$. Now, assuming $\Delta (0) = 0$ the equation (\ref{ddelsine}) can be integrated to yield an explicit solution
\begin{align}
\Delta (t) = \frac{2}{m} \tan^{-1} \left[ \frac{\sqrt{v_0^2 - \kappa^2} \D  \tan \left( \tan^{-1} \left[ \frac{\kappa}{\sqrt{v_0^2 - \kappa^2}} \right] + \frac{mt}{2} \sqrt{v_0^2 - \kappa^2} \right) - \kappa}{v_0} \right]. \label{deltan}
\end{align}
There is a partition in $(\kappa, v_0)$ parameter space given by the equation $|\kappa| = |v_0|$. When $|\kappa|> |v_0|$, so the arguments $\kappa^2 - v_0^2$ of the square roots in (\ref{deltan}) are positive, then there is a family of fixed points of the equation (\ref{ddelsine}), so that the bump position $\Delta (t)$ will eventually become pinned to a single position. In this case, velocity inputs are not detectable by the network, since they do not result in the propagation of a bump. Consistent with this, the equation (\ref{deltan}) has a defined limit at $t \to \infty$, $\Delta (t) \to \frac{2}{m} \tan^{-1} \left[ \sqrt{\kappa^2-v_0^2}/v_0 - \kappa/ v_0 \right]$. The general formulas for all equilibria of (\ref{ddelsine}) are given by
\begin{align}
\bar{\Delta}_{k+} &= \frac{2 k \pi}{m} + \frac{1}{m} \sin^{-1} \frac{v_0}{\kappa} , \hspace{8mm} k=0,...,m-1, \\
\bar{\Delta}_{k-} &=  \frac{(2 k+1) \pi}{m} - \frac{1}{m} \sin^{-1} \frac{v_0}{\kappa}, \hspace{8mm} k=0,...,m-1.
\end{align}
On the other hand, when $|\kappa|< |v_0|$, the heterogeneity $F( \Delta)$ will not lead to pinning of bumps, so bump will propagate indefinitely in response to velocity inputs. However, the heterogeneity will ultimately reduce the speed of propagation of bumps, as found in previous studies of periodically heterogeneous neural fields \citep{bressloff01,kilpatrick08,coombes11}. We can determine the average reduction in the bump's speed by calculating the time $T$ at which $\Delta (t) $ crosses $\Delta = 2 \pi /m$, completing one period of the heterogeneity $\sin (m \Delta)$:
\begin{align}
T = \frac{2 \pi}{m \sqrt{v_0^2 - \kappa^2}},
\end{align}
which means that the average speed of the bump $v_b$ is given
\begin{align}
v_b = \frac{2 \pi/m}{T} = \sqrt{v_0^2 - \kappa^2}, \label{bspeed}
\end{align}
similar to the speed scaling formulas found in \cite{bressloff01,coombes11}. This allows us to directly compute the curve in parameter space at which wave propagation failure occurs, $|v_0| = |\kappa|$ as stated above. Note that the bump speed (\ref{bspeed}) depends on the frequency and amplitude of the heterogeneity through the term
\begin{align}
\kappa = \sigma \frac{\cot (a) \sin (ma) - m \cos (ma)}{m^2 - 1}.
\end{align}
Thus, we can approximate coarse time-average error in path integration for this network as
\begin{align}
r(t) = \Delta_T(t) - \Delta (t) \approx \left( v_0 - \sqrt{v_0^2 - \kappa^2} \right) t.
\end{align}
A more precise estimate is obtained by using the formula for $\Delta (t)$ given by (\ref{deltan}). We demonstrate the accuracy of this full approximation in Fig. \ref{fig4}{\bf C},{\bf D}. Bump position approximations (\ref{deltan}) and average speed approximation computed from (\ref{bspeed}) are compared with the full neural field model (\ref{nfmodel}) are given in Fig. \ref{fig5}. We find the low-dimensional approximation (\ref{ddelsine}) is in excellent agreement with simulations. In particular, the points in parameter space at which propagation failure occur are well matched, and the sinuous trajectory of the bump is well tracked by our low dimensional theory. This suggests we can gain many insights concerning the full model by analyzing this simpler equation (\ref{ddel2}).

As noted above, the existence of spatial heterogeneities in networks can lead to pinning or a reduction in speed of propagating bumps, which should be accurately tracking velocity-input. However, several previous experiments have suggested that sensory feedback is incorporated into the neural representation of spatial navigation \citep{ulanovsky11,battaglia04,zhang14,hardcastle15}. As discussed in section \ref{model}, we propose a simple external control mechanism that incorporates a comparison of an animal's  current estimate of position with an external sensory cue (Fig. \ref{fig1}{\bf A}). In section \ref{control}, we will demonstrate the improvement in position encoding afforded by sensory feedback control. Furthermore, we will show that there is an optimal weighting and timescale of control feedback when sensory cues appear discretely in space or time.

\section{Incorporating sensory cues with online control}
\label{control}

Recent experimental studies have shown that the presence of sensory landmarks reduces the size of mammalian place fields as compared to the case of no landmarks \citep{aikath14,battaglia04,zhang14}. Interestingly, such a reduction in place field size can occur quite quickly, in response to the temporary presence of sensory information, as show in echolocating bats \citep{ulanovsky11}. Note here, we are referring to sensory information beyond the animal's proprioceptive experience of its own motion. Specifically, we are referring to objects placed along the track of navigation that may provide visual, auditory, or olfactory feedback (Fig. \ref{fig1}{\bf B}). This suggests an online interaction between the sensory system and the path integration system that passes some positional information acquired by sensory cues \citep{tsao13}. We suggest that such a mechanisms could counteract errors in position encoding that could arise due to synaptic heterogeneity \citep{hansel13,itskov11,brody03} or noise \citep{laing01,compte00,burak12}. However, when cues occur discretely in space, tuning the strength of feedback introduces a tradeoff between the immediate benefits of recent cues and the deleterious influence of older irrelevant cues. We explore this in the low-dimensional model (\ref{ddel2}) derived in section \ref{potwell}.

\subsection{Reducing error due to network asymmetry and heterogeneity}
\label{reducehet}

We first examine the case of instantaneous cues and updates, modeled as a continuous update to the position, as described by (\ref{contcontr}). This would be the case in which landmark cues are continuously apparent to an animal, allowing the sensory system to send a constant stream of information to the position encoding network. For the time being, we also ignore the impact of noise, exploring its effect in subsection \ref{addnoise}. Under these assumption, the low-dimensional equation for bump position is 
\begin{align}
\dot{\Delta}(t) = F(\Delta (t)) + \phi + v(t) + \lambda (\Delta_T(t)-\Delta(t)).  \label{heterocontr}
\end{align}
As a simple example of the impact of the control term in (\ref{heterocontr}), we examine the case of no heterogeneity $F( \Delta) \equiv 0$ and non-zero asymmetry $\phi > 0$. In this case, we can analytically calculate the long term trajectory of the error $r(t) = \Delta_T(t) - \Delta(t)$. To do so, we can write down the first order differential equation for the error \citep{slotine1991applied}
\begin{align}
\dot{r}(t) + \lambda r(t) &= \dot{\Delta}_T(t) - \dot{\Delta}(t) + \lambda ( \Delta_T(t) - \Delta (t)) \nonumber \\
&= v(t) - \phi - v(t) - \lambda ( \Delta_T(t) - \Delta (t)) + \lambda ( \Delta_T(t) - \Delta (t)) \nonumber \\
& = - \phi. \label{asymede}
\end{align}
It is straightforward to calculate the solution to the linear differential equation (\ref{asymede}) in the case $r(0) = 0$, finding $r(t) = - \phi / \lambda$. Thus, perfect convergence of the trajectory $\Delta (t)$ to $\Delta_T(t)$ can only be obtained in the limit of infinitely strong control $\lambda \to \infty$. It is also important to note that as long as the control strength is positive $\lambda >0$, the error $r(t)$ will be bounded in the long time limit $t \to \infty$.

We can extend our analysis of the equation (\ref{heterocontr}) to the case of arbitrary heterogeneities using regular perturbation theory. Writing the linear expansion of $\Delta (t) = \Delta_0(t) + \Delta_1(t)/\lambda$ under the assumption $\lambda \gg 1$, we find that the zeroth order equation for $\Delta_0(t)$ is simply given by $\Delta_0(t) = \Delta_T(t)$. Extending to the first order equation in $1/ \lambda$, we find
\begin{align*}
\dot{\Delta}_T(t) = F(\Delta_T(t)) + \phi + v(t) - \Delta_1(t).
\end{align*}
Applying the equation $\dot{\Delta}_T(t) = v(t)$, we thus find that $\Delta_1(t) = F(\Delta_T(t)) + \phi$, which means that the long term error can be approximated by
\begin{align*}
r(t) = F \left( \int_0^t v(s) ds \right) + \phi + {\mc O}(1/\lambda^2)
\end{align*}
to first order in $1/ \lambda$. Thus, as long as $F(\Delta)$ is a bounded function, then the error will remain bounded, reaching a maximum amplitude of ${\rm max}_x | F(x) + \phi|$ \citep{slotine1991applied}.

Thus far, we have considered the case of a continuous flow of sensory information providing an accurate estimate of an animal's position in space. However, in more realistic scenarios, animals receive external sensory information discretely in time via local landmarks \citep{battaglia04,tsao13} or echolocation \citep{ulanovsky11}. Sensory cues that provide a landmark for an animal's present position may be captured periodically in time or more randomly; we account for both forms of sensory cue acquisition. As discussed in our formulation of the model in section \ref{model}, we assume the influence of sensory cues weakens as time elapses from the time at which they were received. This is consistent with recent observations concerning the evolution of place fields in bats as a function of the time since the last echo signal \citep{ulanovsky11}. Thus we consider the following model combing path integration with sensory cues acquired at times $t_k$:
\begin{align}
\dot{\Delta}(t) &= F(\Delta (t)) + \phi + v(t) + v_c(t),  \label{discdel} \\
\dot{v_c}(t) &= -v_c(t)/ \tau + \lambda \sum_{k=1}^{N_c} r (t_k) \delta (t-t_k),  \label{discvc} \\
r (t_k) &=  \Delta_T(t_k) - \Delta (t_k).  \nonumber
\end{align}
Analogous to the continuous control case, the error term $r_k$ computes the instantaneous difference between the true position $\Delta_T(t)$ and the encoded position $\Delta (t)$ at time $t_k$. This is then incorporated into the discretely incremented control term $v_c(t)$ with strength $\lambda$, and the temporal decay of cue influence is determined by the timescale $\tau$. We will demonstrate that for any given $\tau$, there is an optimal strength of feedback that trades off the error reduction of recent cues ($t_k$) with the error increase potentially arising for older cues ($t_1,...,t_{k-1}$). Assuming $v_c(0) = 0$ and treating the pointwise values of $r(t)$ as constant, we can integrate (\ref{discvc}) to yield the piecewise smooth function
\begin{align*}
v_c(t) = \lambda \sum_{k=1}^{N_c} r(t_k)  e^{ - (t - t_k) / \tau } H( t - t_k),
\end{align*}
as we did in section \ref{model} for the full neural field model in (\ref{discexp}). Thus, adjustments in velocity are discretely incremented and then decay over time. Also, note in the limit $\tau \to 0$ and $t_{k+1} - t_k \to 0$, we obtain the continuous control function $v_c(t) = \lambda r(t)$. This can be seen by performing this limit on (\ref{discvc}) and then integrating.

\begin{figure}
\begin{center} \includegraphics[width=5cm]{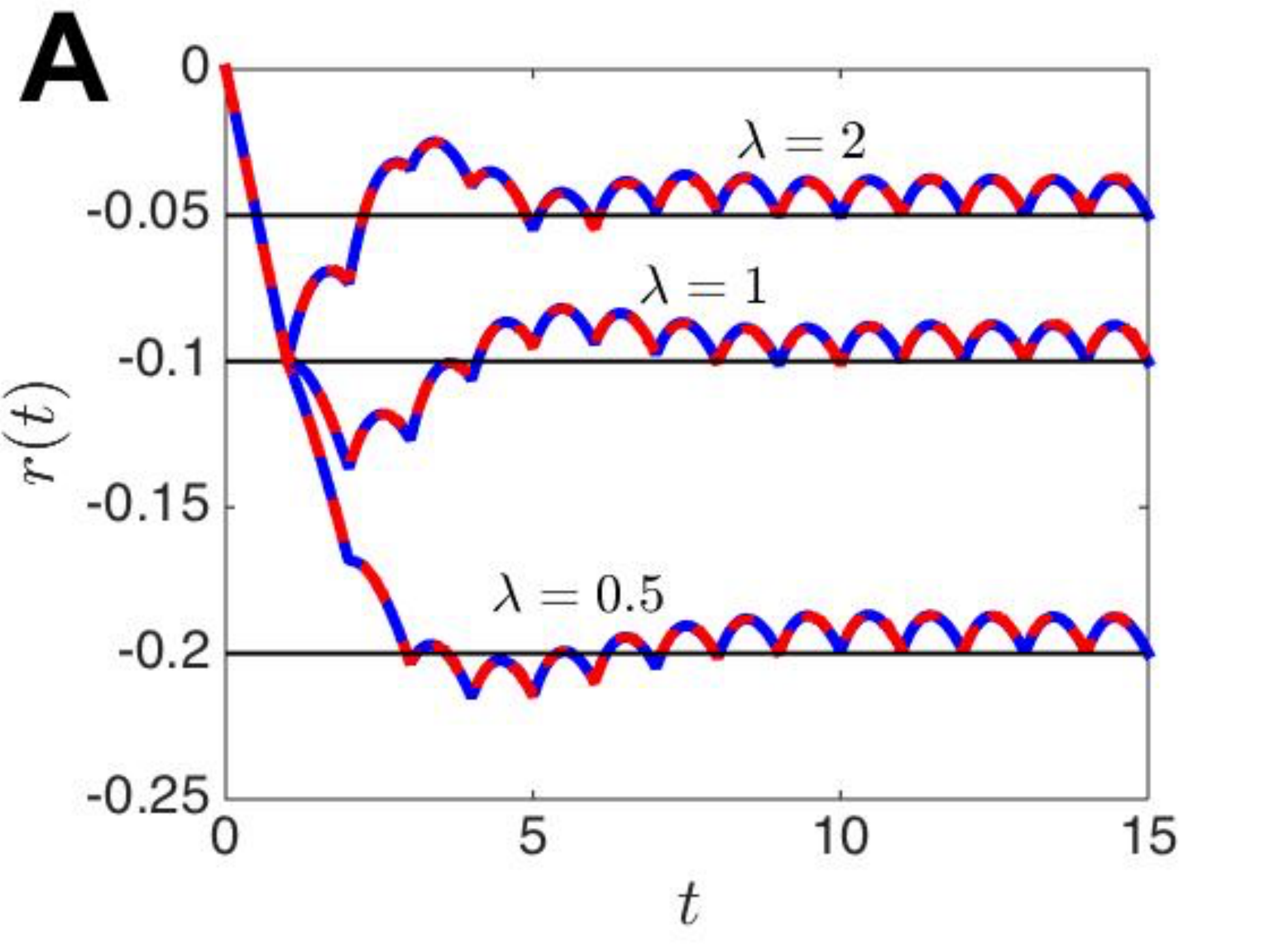} \includegraphics[width=5cm]{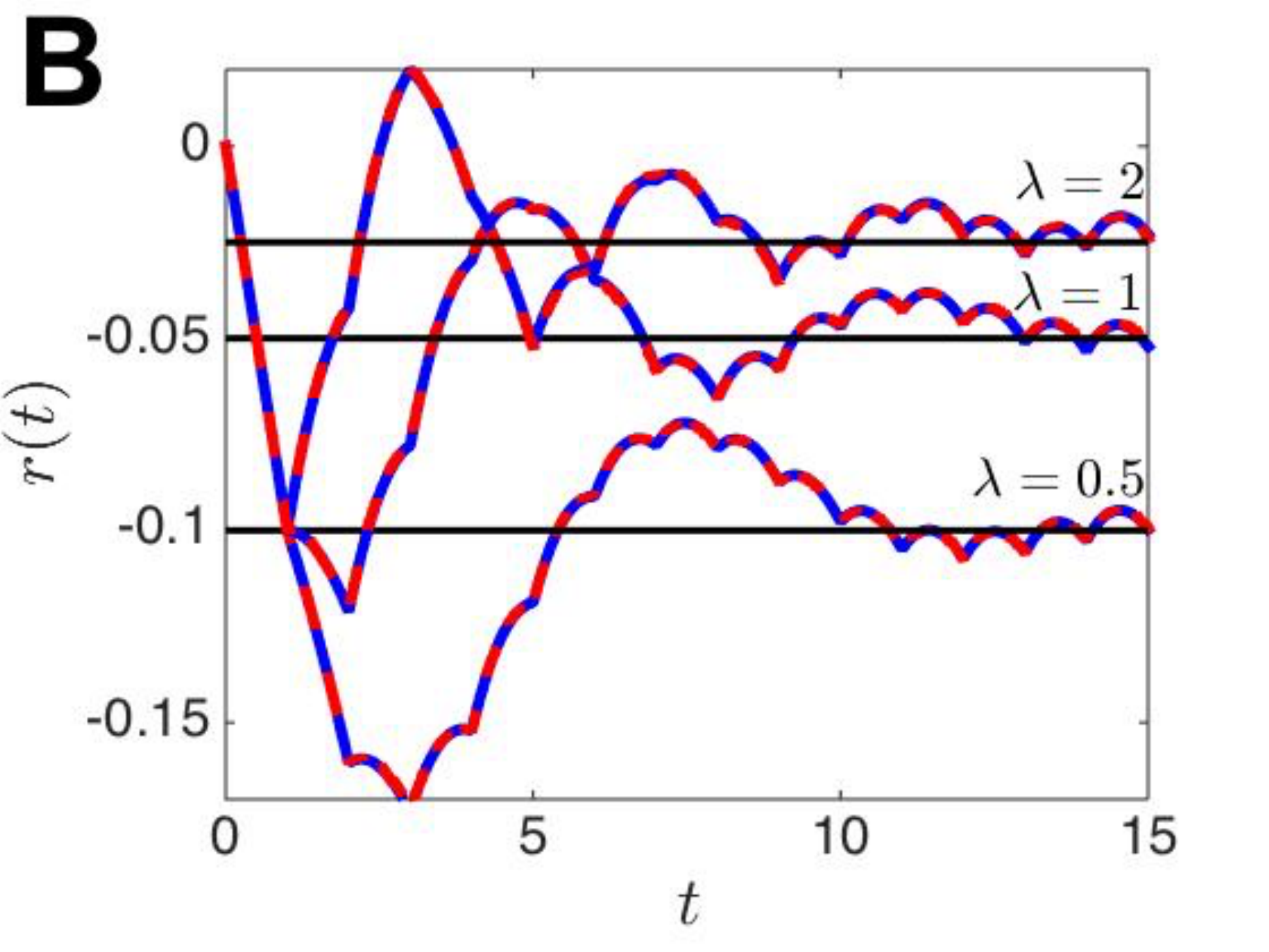} \includegraphics[width=5cm]{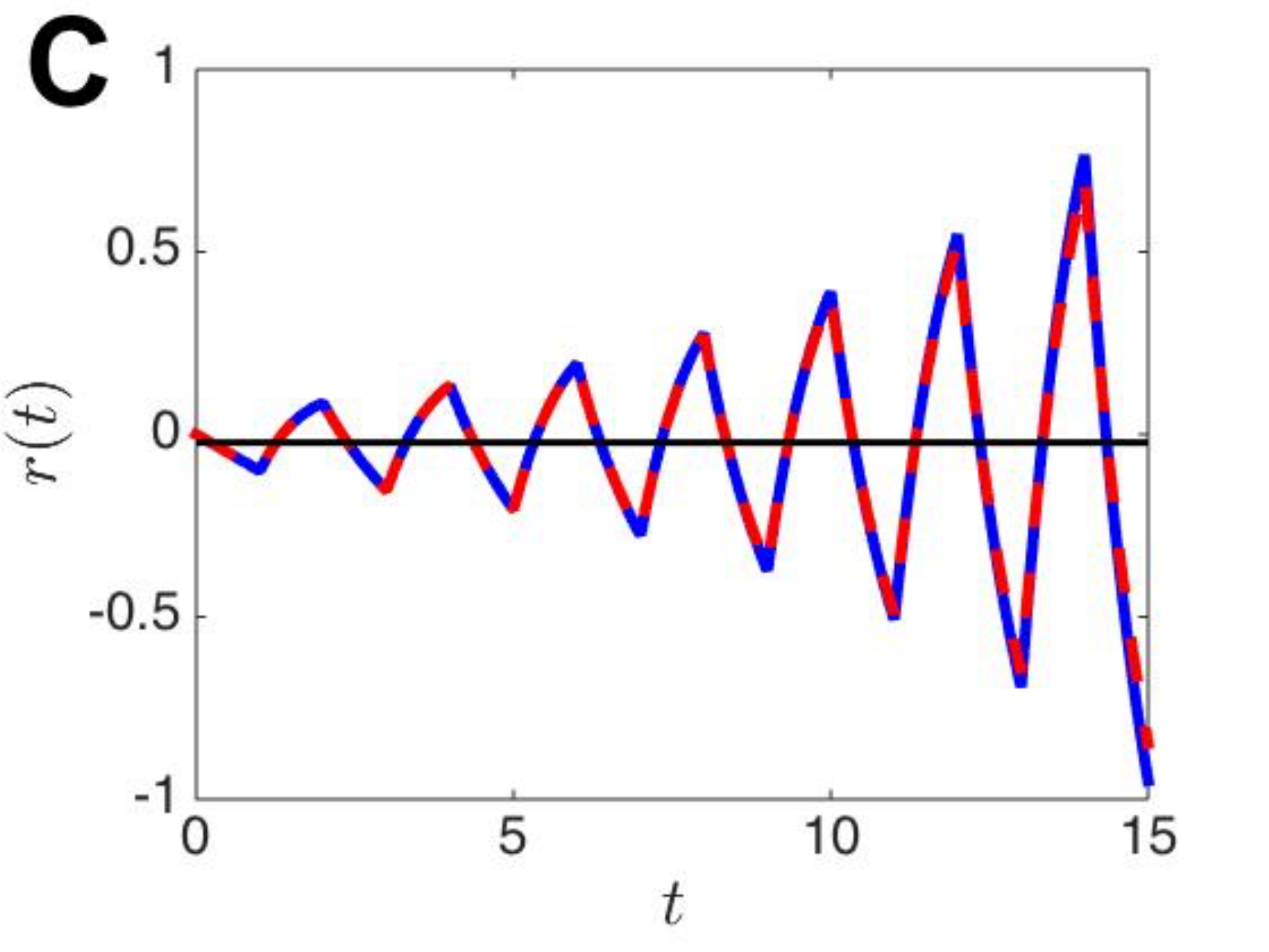} \end{center}
\caption{Path integration error in a network with asymmetry $\phi = 0.1$, discretely controlled according to (\ref{discvc}). ({\bf A}) Error resulting from asymmetry plus discrete control with time decay $\tau = 1$ quickly reaches the fixed point $r^*$ (thin lines) given by (\ref{dasymfp}). Notice as the strength of control $\lambda$ is increased, the long term error decreases. Low dimensional approximation (solid lines) given by (\ref{delaexp}) is in excellent agreement with simulations (dashed lines) of the full model (\ref{nfmodel}). ({\bf B}) Increasing the control decay timescale $\tau = 2$ leads to longer lasting oscillations in the error. ({\bf C}) Making the control too strong, $\lambda = 4.5$, leads to instability in the error. Negative feedback produces oscillations that grow in amplitude. Time spacing between cues is $\Delta t = 1$. Other parameters $\theta = 0.5$, $v_0 = 0.1$. Numerical simulations utilize the same parameters as in Fig. \ref{fig2}.}
\label{fig6}
\end{figure}

To demonstrate the impact of discrete control in more detail, we begin by studying the case of a network subject only to asymmetry ($F(\Delta) \equiv 0$ and $\phi >0$). Furthermore, we focus on the case of constant velocity input $v(t) \equiv v_0$, so we can write the discretely controlled position equation (\ref{discdel}) as
\begin{align}
\dot{\Delta}(t) = \phi + v_0 + \lambda \sum_{k=1}^{N_c} \left[ v_0 t_k - \Delta (t_k) \right] \e^{-(t-t_k)/\tau} H(t - t_k). \label{delaexp}
\end{align}
We can solve the piecewise smooth differential equation (\ref{delaexp}) recursively, integrating with a new initial condition $\Delta (t_k)$ at each cue time $t_k$. In the initial time domain $[0,t_1)$, $\Delta(0) = 0$ and $\dot{\Delta}(t) = v_0 + \phi$, so $\Delta (t) = (v_0 + \phi) t$ and $\Delta (t_1) = (v_0 + \phi) t_1$. On the subsequent time domain $[t_1,t_2)$, we have
\begin{align*}
\dot{\Delta}(t) = v_0 + \phi + r(t_1) \e^{-(t-t_1)/\tau},
\end{align*}
so
\begin{align*}
\Delta (t) = (v_0 + \phi) t + \lambda \tau r(t_1) \left[ 1 - \e^{-(t-t_1)/\tau} \right].
\end{align*}
In a similar way, we can solve for $\Delta (t)$ on $[t_2,t_3)$ to find
\begin{align*}
\Delta (t) = (v_0 + \phi) t + \lambda \tau \sum_{k=1}^{2} r(t_k) \left[ 1 - \e^{-(t-t_k)/\tau} \right],
\end{align*}
and in general, we can express
\begin{align*}
\Delta (t) = (v_0 + \phi )t + \lambda \tau \sum_{k=1}^{N_c} r(t_k) \left[ 1 - \e^{- (t-t_k)/\tau} \right] H( t- t_k).
\end{align*}
Thus, we can express the error as a function of time
\begin{align}
r(t) = - \phi t - \lambda \tau \sum_{k=1}^{N_c} r(t_k) \left[ 1 - \e^{-(t-t_k)/ \tau} \right] H(t- t_k). \label{discerr1}
\end{align}
Expressing $r_k := r(t_k)$ and focusing on the error at the cue timepoints $t_k$, we can write (\ref{discerr1}) as
\begin{align*}
r_l = - \phi t - \lambda \tau \sum_{k=1}^{l-1} r_k \left[ 1 - \e^{-(t_l-t_k)/ \tau} \right]. 
\end{align*}
Furthermore, in the case of periodically spaced cues, we can write $t_{k+1} - t_k =  (\Delta t)$, $\forall k$, so that
\begin{align}
r_l = - \phi \cdot l \cdot \Delta t - \lambda \tau \sum_{k=1}^{l-1} r_k \left[ 1 - \e^{- (l-k) \Delta t/ \tau} \right]. \label{discerr3}
\end{align}
Assuming that $\lambda$ is not too large, the discrete equation (\ref{discerr3}) will have a fixed point in the long time limit $r_l \to r^*$, which we can compute by taking the difference between $r_{l+1}$ and $r_{l}$ and approximating $r_k \approx r^*$:
\begin{align*}
r_l = r^* &= - \phi \cdot l \cdot \Delta t - \lambda \tau r^* \sum_{k=1}^{l-1} \left[ 1 - \e^{-(l-k) \Delta t/ \tau} \right], \\
r_{l+1} = r^* &= - \phi \cdot (l+1) \cdot \Delta t - \lambda \tau r^* \sum_{k=1}^{l} \left[ 1 - \e^{-(l+1-k) \Delta t/ \tau} \right],
\end{align*}
and we can make the approximation $\e^{-l \Delta t/ \tau} \to 0$, so that $r_l - r_{l+1}$ yields
\begin{align}
0 &= \phi \cdot \Delta t + \lambda \tau r^* \hspace{4mm} \Rightarrow \hspace{4mm} r^* = - \phi \cdot \Delta t/ (\lambda \tau).  \label{dasymfp}
\end{align}
We demonstrate the accuracy of the formula in Fig. \ref{fig6}{\bf A},{\bf B}, showing that the error remains bounded due to the periodic perturbations of the discrete control term. Of course, the fixed point value given by (\ref{dasymfp}) is contingent on its existence and stability. In cases where either condition is violated, the error $r_l$ will diverge in the long time limit (Fig. \ref{fig6}{\bf C}). Essentially, negative feedback overcorrects for the previously observed errors at each cue time, leading to unstable oscillations in the error. Analytically identifying the cases in which $r_l$ diverges would require a more thorough study of the discrete equation (\ref{discerr3}). Numerical simulations suggest there is a boundary value of $\lambda$ above which these unstable oscillations occur. Thus, the maximal value $\lambda$ for which the fixed point $r^*$ exists and is stable would correspond to the optimal control strength, all other parameters being fixed.

We now study the case of heterogeneities and explore the impact of sensory cues on the long term error. Note, in the case of no asymmetry and constant velocity input $v(t) = v_0$, the low dimensional equation for bump position takes the form
\begin{align}
\dot{\Delta}(t) = F(\Delta(t)) + v_0 + \lambda \sum_{k=1}^{N_c} \left[ v_0 t_k - \Delta (t_k) \right] \e^{-(t- t_k)/ \tau} H(t- t_k).  \label{dischet}
\end{align}
While we cannot solve (\ref{dischet}) explicitly for general heterogeneities $F(\Delta)$, we can numerically analyze the impact of both the control strength $\lambda$ and the control decay timescale $\tau$ on the long term error $r(t) = \Delta_T(t) - \Delta (t)$. Specifically, we associate error with a scalar quantity by computing the log of the $L^2$-norm
\begin{align}
R := \ln || \Delta_T(t) - \Delta(t)||_2 = \ln \left[ \sqrt{\int_0^{t_f} | \Delta_T(t) - \Delta (t)|^2 dt} \right],  \label{logerr}
\end{align}
where $t_f$ is time at which the path ends. We compare the effects of varying the spacings $t_{k+1} - t_k$ between subsequent cues, testing both time-periodic cues ($t_{k+1} - t_k = \Delta t$, $\forall k$) and exponentially distributed spacings ($p( \Delta t) = \mu \e^{- \mu \Delta t}$). Furthermore, we randomize the heterogeneity according to the formula (\ref{ewf}) with variance $\sigma_n^2 = 1$ with four total modes ($N=2$). To average error across many realizations, we simulate the controlled version equation (\ref{dischet}) for many different randomly generated heterogeneities, compute an $L^2$-norm of error $R_j$ for the $j$th realization and average $\langle R \rangle = \frac{1}{N_r} \sum_{j=1}^{N_r} R_j$ for $N_r$ realizations. 

\begin{figure}
\begin{center} \includegraphics[width=7cm]{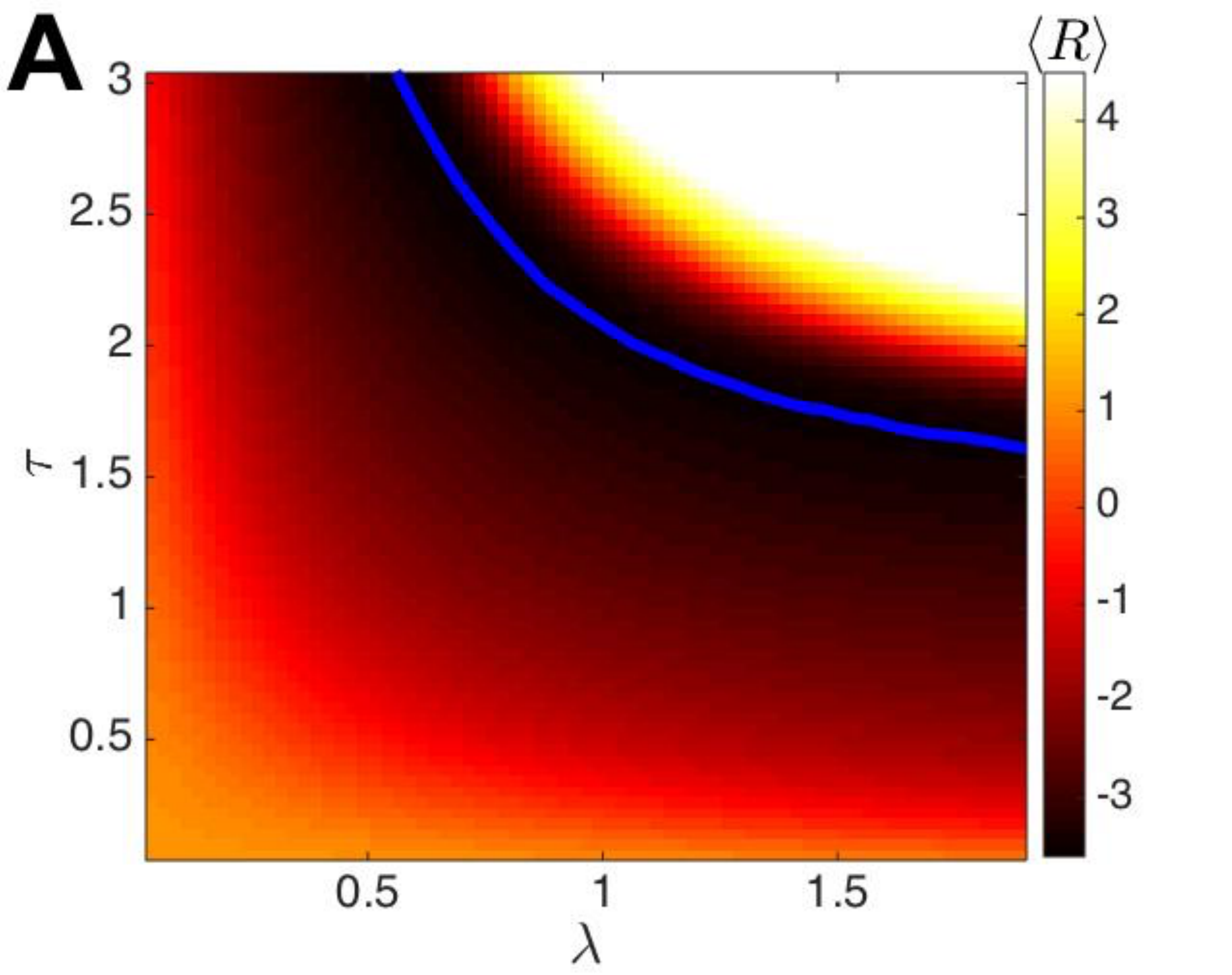}
\includegraphics[width=7cm]{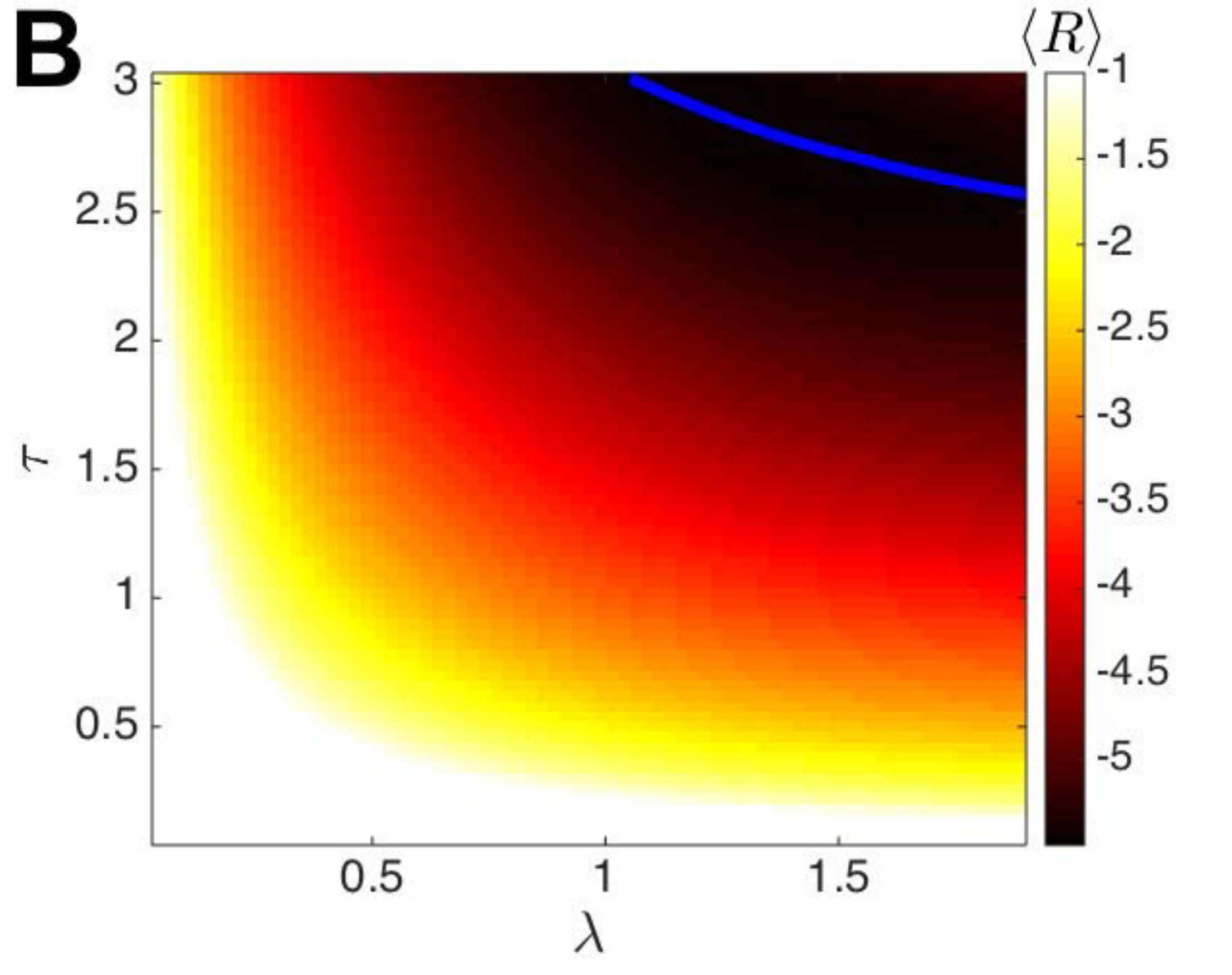} \\
\includegraphics[width=7cm]{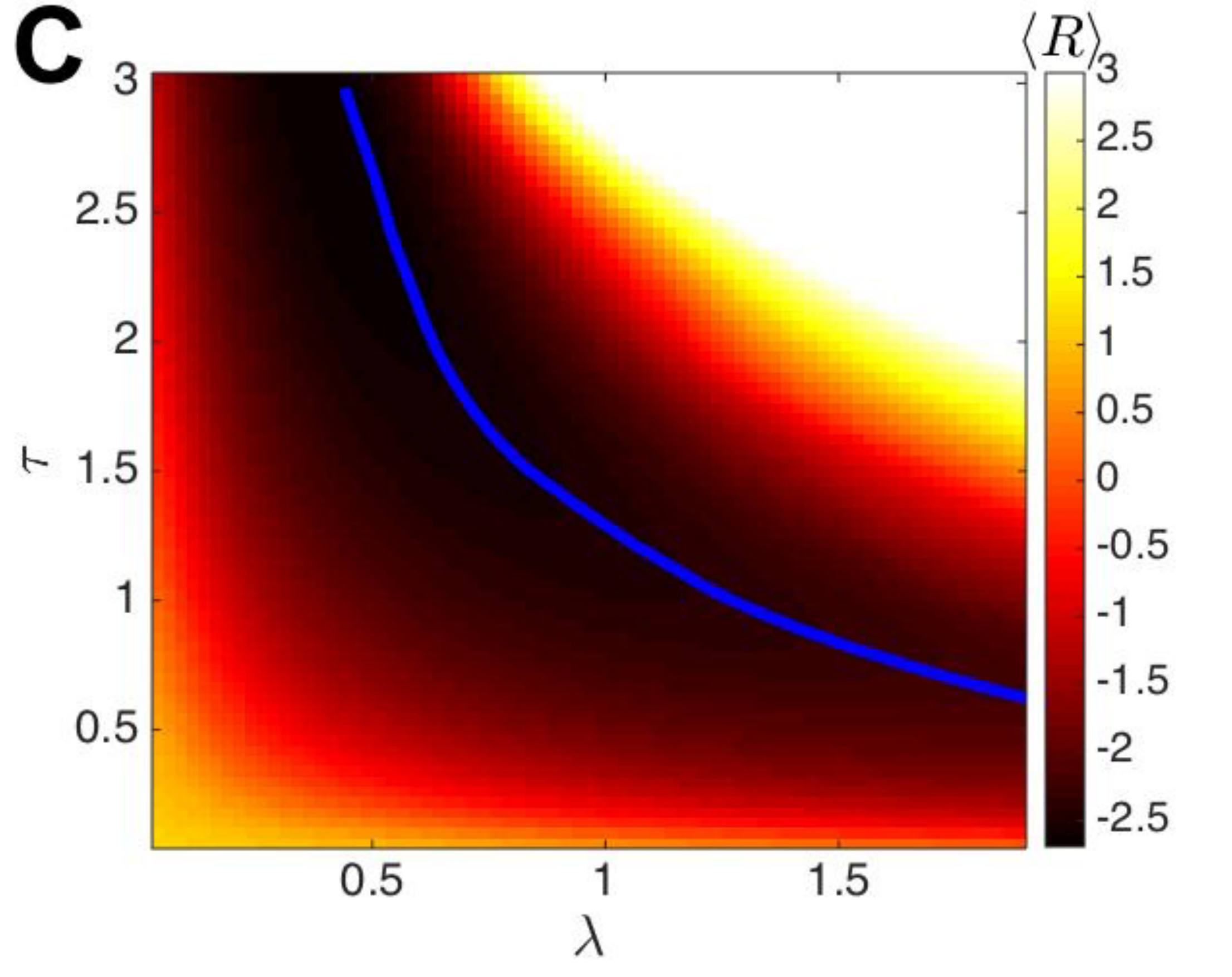}
\includegraphics[width=7cm]{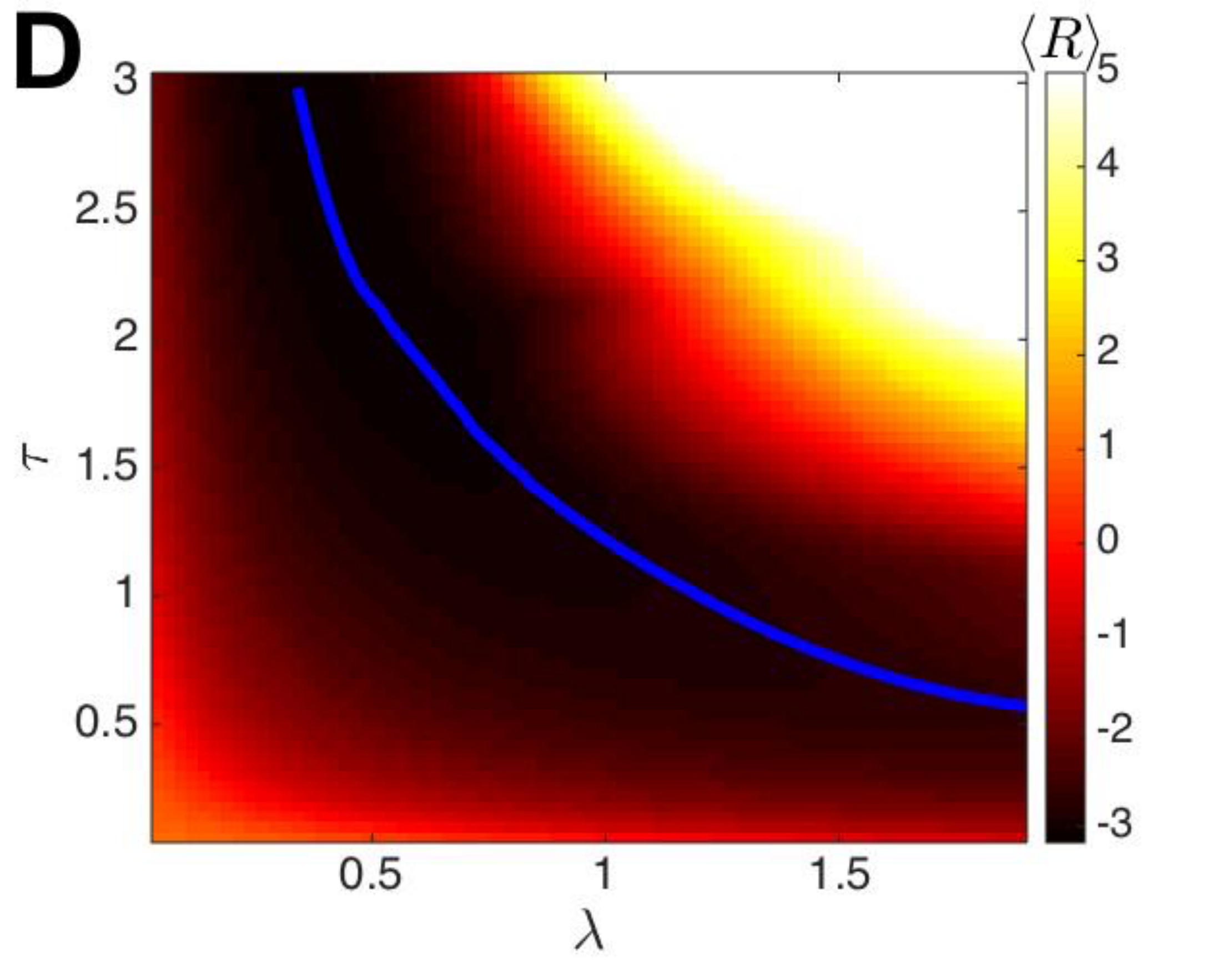} \end{center}
\caption{Average log error $\langle R \rangle$ computed across realizations of (\ref{logerr}) for the discretely controlled low-dimensional approximation (\ref{dischet}) with spatial heterogeneity resulting from (\ref{ewf}) with $N=2$ and coefficient variance $\sigma_n^2 = 1$. ({\bf A}) For periodically appearing control cues with $\Delta t = 4$, we find there is an intermediate curve (solid) of $(\lambda, \tau)$ values that minimizes $\langle R \rangle$. In particular, as the control decay timescale $\tau$ is increased, the optimal value of $\lambda$ decreases. ({\bf B}) The same trend is consistent for periodic cues with spacing $\Delta t = 2$, but the curve of optimal $(\lambda, \tau)$ values shifts so there are higher values of $\tau$ associated with each value of $\lambda$ as compared with {\bf A}. ({\bf C},{\bf D}) When spacings between cue times are exponentially distributed $p( \Delta t) = \mu \e^{- \mu \Delta t}$ with $\mu = 0.5$ in {\bf C} and $\mu = 1$ in {\bf D}, we find the optimal curve shifts to shorter values of $\tau$ for more frequent cues. Other parameters $v_0 = 0.15$, $\theta = 0$, $\sigma = 0.1$, and simulation time $t_f=40$. Numerical simulations are performed using Euler's method with a timestep of 0.05, and each grid point used 1000 realizations.}
\label{fig7}
\end{figure}

We are mainly interested in the $(\lambda, \tau)$ values that minimize the average log error $\langle R \rangle$. Our findings are summarized in Fig. \ref{fig7}. First, we note that there is always a curve through $(\lambda, \tau)$ space determining the values of the control term that minimize the average error $\langle R \rangle$. In all plots, the associated $\tau$ value decreases with $\lambda$ and vice versa. In general, we find this relationship to be roughly inversely proportional $\lambda \propto 1/ \tau$. This means that stronger control should decay more quickly, and equivalently weaker control can last longer. Furthermore, by comparing plots for periodic cues with $\Delta t = 4$ (Fig. \ref{fig7}{\bf A}) versus $\Delta t = 2$ (Fig. \ref{fig7}{\bf B}), we find longer decay timescales associated with each $\lambda$ value in the case $\Delta t =2$. Such a trend may arise due to the fact that more frequent updates in sensory information via cues prevents overcorrection that could occur in the case of less frequent curves. A similar trend arises in the case of exponentially distributed time spacings between cues ($\mu = 0.5$ in Fig. \ref{fig7}{\bf C} versus $\mu = 1$ in Fig. \ref{fig7}{\bf D}). When cues are more frequent, the optimal timescale of decay $\tau$ is slightly larger for each value of $\lambda$.

\begin{figure}
\begin{center} \includegraphics[width=7.5cm]{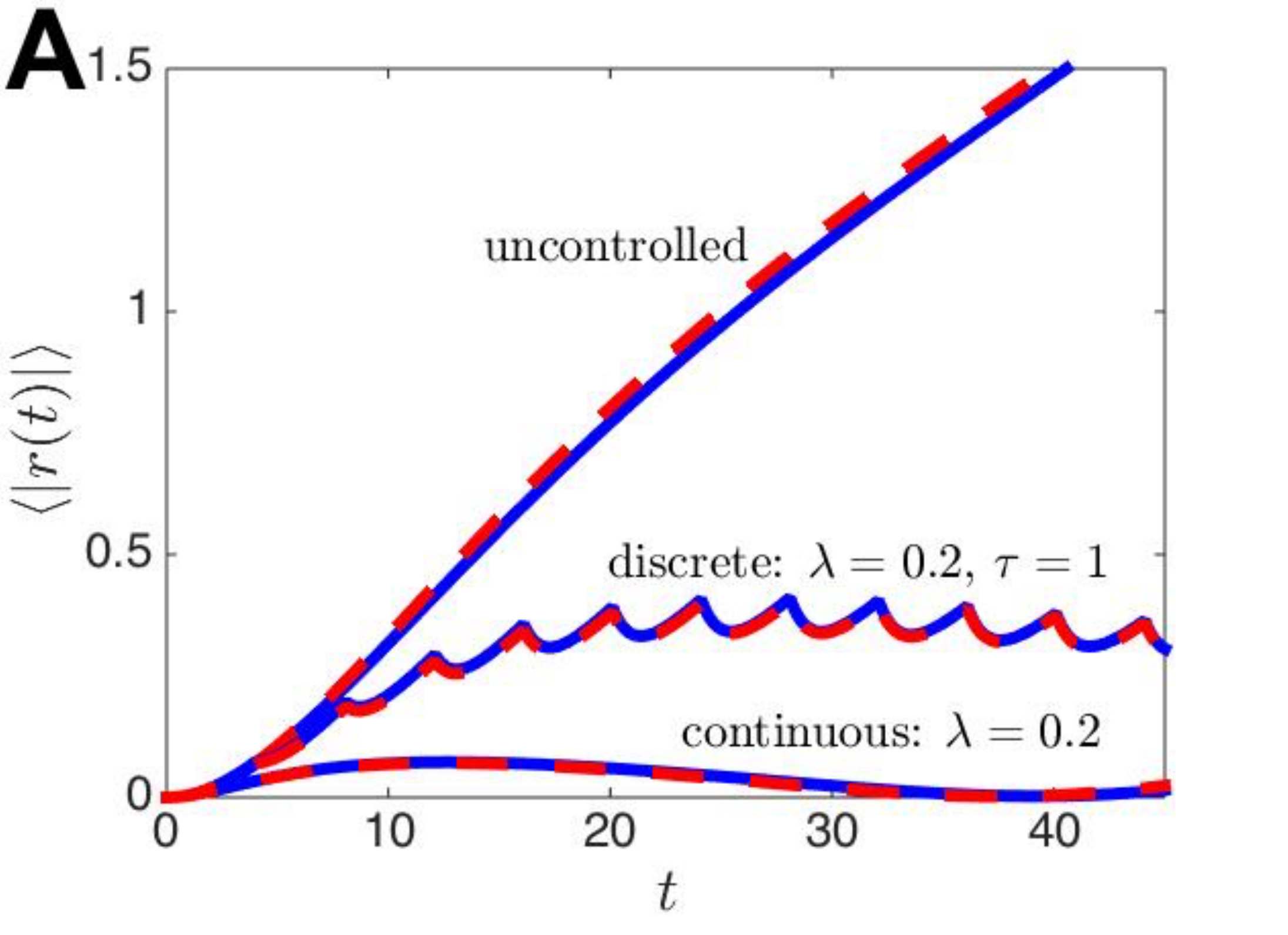} \includegraphics[width=7.5cm]{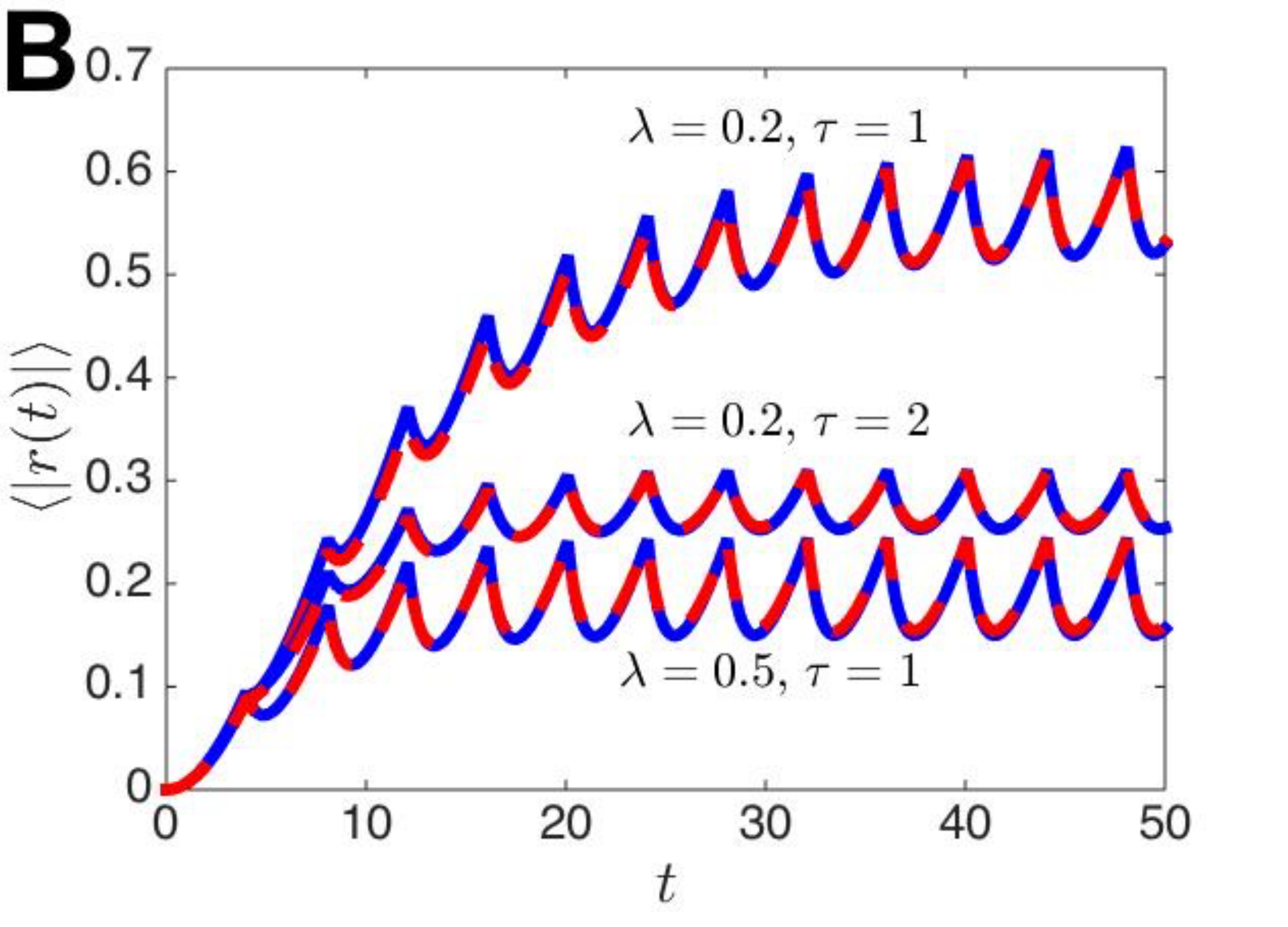} \end{center}
\caption{Average error $\langle |r(t)| \rangle$ as a function of time in a heterogeneous network ($w_u(x) = \alpha_1 \cos (x)$ with control. Velocity-input is constant $v(t) = v_0 = 0.05$. ({\bf A}) Numerical simulations (dashed lines) of the neural field (\ref{nfmodel}) are well matched by the low-dimensional approximation (solid lines) given by (\ref{dischet}). As demonstrated, continuous control provides the best reduction in error, but discrete control with $\Delta t = 4$ still provides an appreciable reduction. ({\bf B}) Plot demonstrates the impact of varying the control strength $\lambda$ and the control decay timescale $\tau$. Other parameters are $\theta = 0.2$ and $\sigma = 0.1$. Numerical simulations are run with the same parameters as in Fig. \ref{fig2} for 1000 realizations each curve.}
\label{fig8}
\end{figure}

In addition, we have studied the average error as a function of time for both continuously and discretely controlled networks with heterogeneities. Note that we randomize the heterogeneity $w_u(x) = \alpha_1 \cos (x)$ so that $\alpha_1$ is normally distributed with variance unity. To compute the average error, we take the mean of the absolute value $\langle |r(t) | \rangle$ as shown in Fig. \ref{fig8}. In the case of strong continuous control, it is possible to substantially decrease the impact of heterogeneities as compared with the uncontrolled case (Fig. \ref{fig8}{\bf A}). Discrete control maintains an intermediate level of error, since there is not a constant stream of information provided to reduce error. Varying the strength $\lambda$ and timescale $\tau$ of control alters the long term variance in the error (Fig. \ref{fig8}{\bf B}). As suggested by Fig. \ref{fig7}, strong and fast decaying control tends to lead to substantial reductions in error.

\subsection{Reducing error due to dynamic fluctuations} 
\label{addnoise}

We now examine the impact of sensory feedback on networks subject to temporal noise fluctuations. Dynamic variability in networks can arise from ion channel fluctuations \citep{chow96}, synaptic variability \citep{ribrault11}, or finite size effects \citep{bressloff09}. As demonstrated in our analysis in section \ref{potwell}, we can derive a reduced equation for the position of a velocity-driven bump subject to noise to a single stochastic differential equation (\ref{ddel2}). Focusing specifically on the impact of noise, taking constant speed $v(t) \equiv v_0$, and ignoring heterogeneities, we find that the controlled equation for the bump position takes the form
\begin{align}
d \Delta (t) = \left[ v_0 + v_c(t) \right] dt + d {\mc W}(t).  \label{noisecontr}
\end{align}

\begin{figure}
\begin{center} \includegraphics[width=7.3cm]{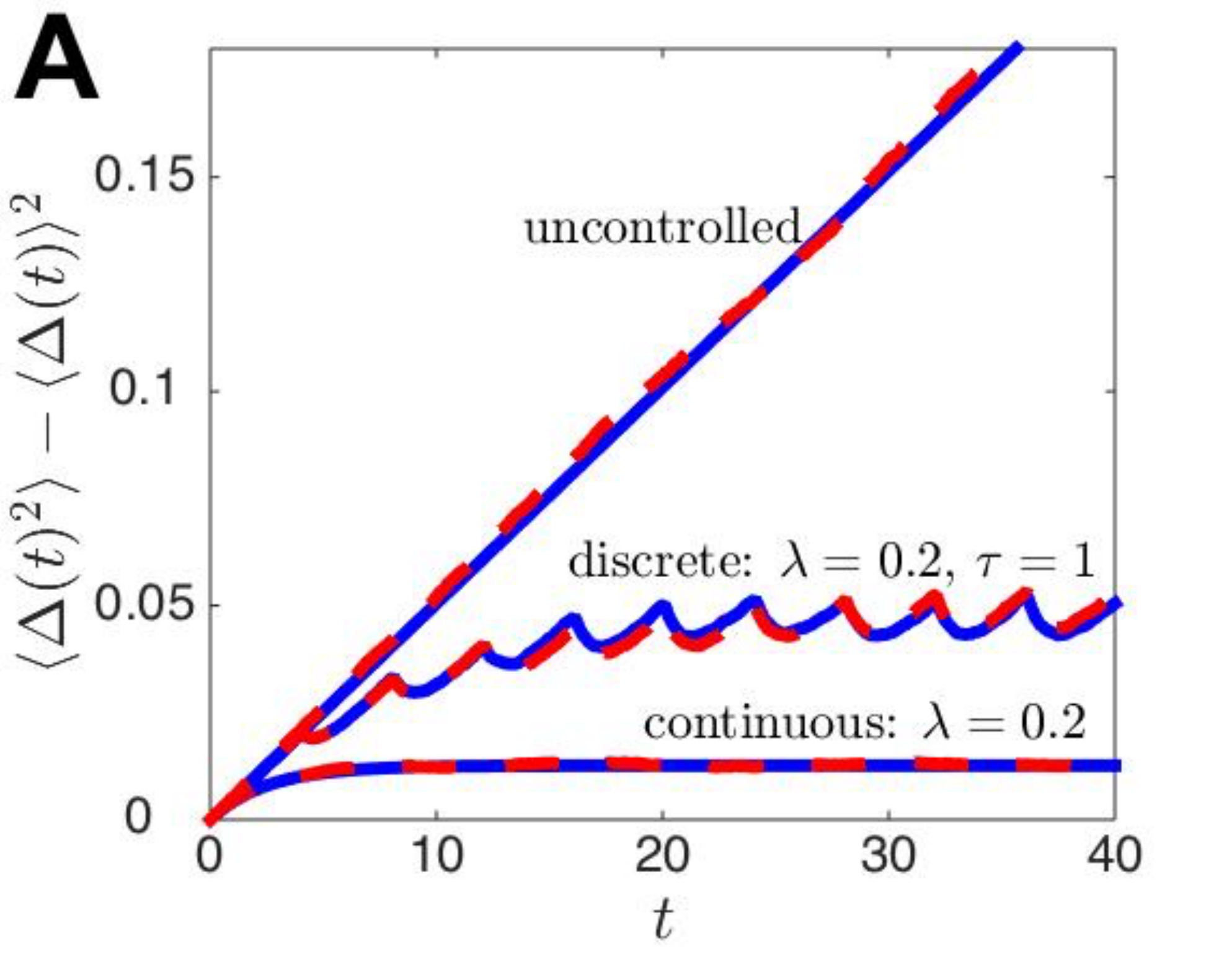} \includegraphics[width=7.7cm]{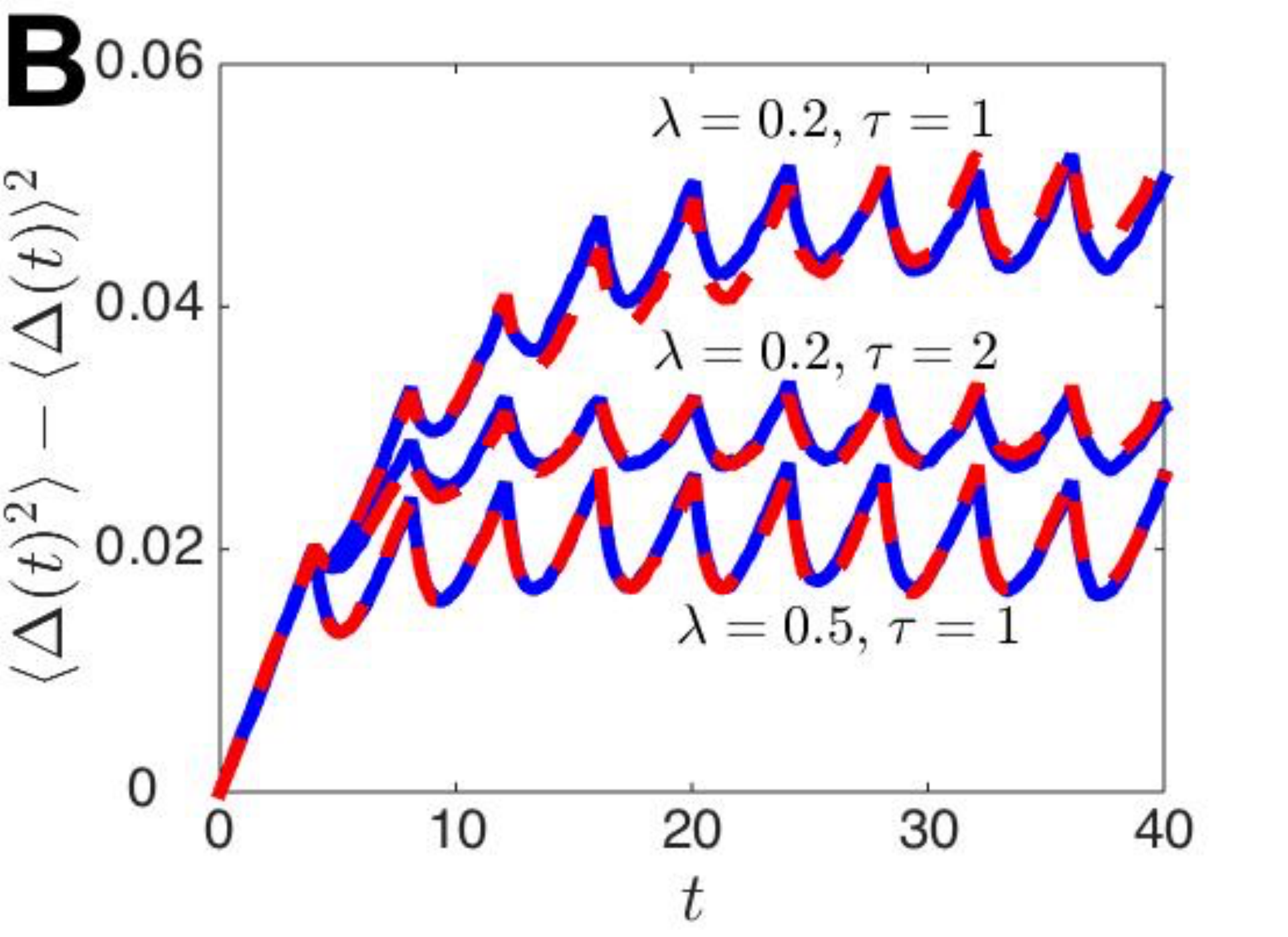} \end{center}
\caption{Variance $\langle \Delta (t)^2 \rangle - \langle \Delta (t) \rangle^2$ computed for the noise-driven network with control. Velocity-input is constant $v(t) = v_0 = 0$. ({\bf A}) Numerical simulations (dashed lines) of the neural field model (\ref{nfmodel}) are well matched to low-dimensional approximation (solid lines) given by (\ref{noisecontr}). Continuous control substantially reduces the variance, but discrete control with $\Delta t = 4$ also provide a variance reduction. Notably, the variance saturates in the case of discrete control as well. ({\bf B}) Similar to the case of quenched variability through heterogeneity in Fig. \ref{fig8}, varying the strength and timescale of control alters the long term variance. Other parameters are $\theta = 0.2$ and $\epsilon = 0.1$. Numerical simulations are run with the same parameters as in Fig. \ref{fig2} with 1000 realizations each curve.}
\label{fig9}
\end{figure}

We begin by examining the case of continuous sensory feedback, in which case (\ref{noisecontr}) becomes
\begin{align}
d \Delta (t) = \left[ v_0 + \lambda v_0 t - \lambda \Delta (t) \right] dt + d {\mc W}(t).  \label{noisecont}
\end{align}
Note that (\ref{noisecont}) is a non-autonomous Ornstein-Uhlenbeck process, and we can use integrating factors to identify an explicit solution. Utilizing the change of variables $h( \Delta, t) = \Delta \e^{\lambda t}$ and differentiating with respect to $t$, we find
\begin{align}
d h( \Delta, t) &= d \Delta \e^{\lambda t} + \lambda \Delta \e^{\lambda t} = \e^{\lambda t} \left[  v_0 + \lambda v_0 t - \lambda \Delta + \lambda \Delta \right] dt + \e^{\lambda t} d {\mc W} \nonumber \\
&= \e^{\lambda t} \left[ v_0 + \lambda v_0 t \right] dt + \e^{\lambda t} d {\mc W} = d \left( v_0 t \e^{\lambda t} \right) + \e^{\lambda t} d {\mc W}. \label{intfact}
\end{align}
Assuming $\Delta (0) = 0$, we can integrate (\ref{intfact}) and multiply through by $\e^{-\lambda t}$ to yield the solution
\begin{align*}
\Delta (t) = v_0 t + \e^{- \lambda t} \int_0^t \e^{\lambda s} d {\mc W}(s),
\end{align*}
whose mean is $\langle \Delta (t) \rangle = v_0 t$ and variance is given
\begin{align*}
\langle \Delta (t)^2 \rangle - \langle \Delta (t) \rangle^2 = \frac{\sqrt{D}}{2 \lambda} \left[ 1 - \e^{- 2 \lambda t} \right],
\end{align*}
where the diffusion coefficient $D$ can be calculated from the neural field model parameters as in (\ref{dcoef}). The long term variance is thus given by $\D \lim_{t \to \infty} \langle \Delta (t)^2 \rangle - \langle \Delta (t) \rangle^2 = \frac{\sqrt{D}}{2 \lambda}$. Note that in the limit $\lambda \to \infty$, the variance goes to zero $\langle \Delta (t)^2 \rangle - \langle \Delta (t) \rangle^2 \to 0$, suggesting that strengthening continuous control will always reduce the average error further. Continuous control substantially reduces the long term variance in the bump position $\Delta (t)$ as well as the error
\begin{align*}
r(t) = \Delta_T(t) - \Delta (t) = - \e^{- \lambda t} \int_0^{t} \e^{\lambda s} d {\mc W}(s).
\end{align*}
Note that $\langle r(t) \rangle = 0$ and $\langle r(t)^2 \rangle = \frac{\sqrt{D}}{2 \lambda} \left[ 1 - \e^{- 2 \lambda t} \right]$. We compare the continuously controlled system to the control-free system in Fig. \ref{fig9}{\bf A}, revealing the long term saturation in the position variance.

We also study the effect of discrete control on the variance in position, using the low-dimensional approximation of bump position
\begin{align}
d \Delta (t) &= \left[ v(t) + v_c(t) \right] dt + d {\mc W}(t), \label{noisedcont} \\
\dot{v_c}(t) &= - \frac{v_c}{\tau} + \lambda \sum_{k=1}^{N_c} r(t_k) \delta (t- t_k) .\nonumber
\end{align}
Again, this is under the assumption that cues are spaced discretely in time or space, and they provide sensory input for a brief period of time. As in subsection \ref{reducehet}, we can solve (\ref{noisedcont}) iteratively. To begin, note that when $t \in [0,t_1)$, $\Delta (t)$ has yet to be affected by the feedback control term in (\ref{noisedcont}), so $\Delta (t) = v_0 t + {\mc W}(t)$. Subsequently, we can integrate (\ref{noisedcont}) to find the stochastic formula for $\Delta (t)$ after the first cue at $t_1$:
\begin{align*}
\Delta (t) = v_0 t + \lambda \tau ( v_0 t_1 - \Delta (t_1)) \left[ 1 - \e^{- (t-t_1)/\tau} \right] + {\mc W} (t), \hspace{4mm} t \in [t_1,t_2).
\end{align*}
Each realization will typically produce a different value for $\Delta (t_1) = v_0 t_1 + {\mc W}(t_1)$. Note that $\langle \Delta (t) \rangle = v_0 t$, so the impact of noise and control can be observed by calculating the variance \citep{gardiner04}
\begin{align}
\langle \Delta(t)^2 \rangle - \langle \Delta(t) \rangle^2 =&  \left\langle \left( v_0t + \lambda \tau (v_0 t_1 - \Delta(t_1)) \left( 1 - e^{- (t - t_1) / \tau } \right) + {\mc W}(t) \right)^2 \right\rangle
- (v_0t)^2 \nonumber  \\
=& \langle {\mc W}(t)^2 \rangle + \left\langle (\lambda \tau )^2 {\mc W}(t_1)^2 \left( 1 - \e^{-(t-t_1)/\tau} \right)^2 \right\rangle - \left\langle 2 \lambda \tau {\mc W}(t_1) {\mc W}(t)  \left( 1 - \e^{-(t-t_1)/\tau} \right) \right\rangle \nonumber \\
=&   D t + \lambda \tau D t_1 \left( 1-\e^{-(t-t_1)/\tau} \right) \left( \lambda \tau \left( 1 - \e^{-(t-t_1)/\tau} \right) - 2 \right). \label{noisevar}
\end{align}
One insight to be gained from the formula (\ref{noisevar}) is that infinitely strong and fast decaying control, even when it is discrete, will minimize the variance in the position. Specifically, if we take $\lambda = \lambda_0/ \tau$, then we can write
\begin{align*}
\langle \Delta(t)^2 \rangle - \langle \Delta(t) \rangle^2 =&   D t + \lambda_0 D t_1 \left( 1-\e^{- \lambda (t-t_1)/\lambda_0} \right) \left( \lambda_0 \left( 1 - \e^{- \lambda (t-t_1)/\lambda_0} \right) - 2 \right).
\end{align*}
Taking the limit as $\lambda \to \infty$, we find that
\begin{align*}
\lim_{\lambda \to \infty} \left[ \langle \Delta(t)^2 \rangle - \langle \Delta(t) \rangle^2 \right] =&   D t + \lambda_0 D t_1  ( \lambda_0 - 2 ),
\end{align*}
which is minimized when the scaling term $\lambda_0 = 1$, yielding $\langle \Delta(t)^2 \rangle - \langle \Delta(t) \rangle^2  = D(t-t_1)$.

We can solve (\ref{noisedcont}) explicitly for an arbitrary number of cue times, yielding
\begin{align}
\Delta (t) = v_0 t + \lambda \tau \sum_{k=1}^{N_c} r(t_k) \left[ 1 - \e^{- (t-t_k)/\tau} \right] H(t- t_k) + {\mc W}(t).  \label{noisepos}
\end{align}
While it is clear that the mean $\langle \Delta (t) \rangle = v_0 t$, it is more complicated to compute the variance $\langle \Delta (t)^2 \rangle - \langle \Delta (t) \rangle^2 $ in general. This is chiefly due to the fact that $r(t_k)$ will depend on $(r(t_1),...,r(t_{k-1}))$, and this long-lasting history-dependence will accumulate indefinitely. To gain some analytical understanding, we make the assumption of brief control impulses, so that $\tau \ll 1$ and $\e^{-(t_{k+1} - t_k)/\tau} \ll 1$, $\forall k$. In this case, we can write the equation for the update of the error term $r_k := r(t_k)$ as
\begin{align*}
r_{l+1} \approx (1 - \lambda \tau) r_k + {\mc W}(t_{k+1}) - {\mc W}(t_k),
\end{align*} 
where $r_1 = {\mc W}(t_1)$. Again, it should be clear there is an inverse relationship between the impact of $\lambda$ and $\tau$ on the long term error in this limiting case. We compute the variance numerically from (\ref{noisepos}) for the case of discrete control in Fig. \ref{fig9}, demonstrating an excellent match with the neural field model (\ref{nfmodel}).

\begin{figure}
\includegraphics[width=7.7cm]{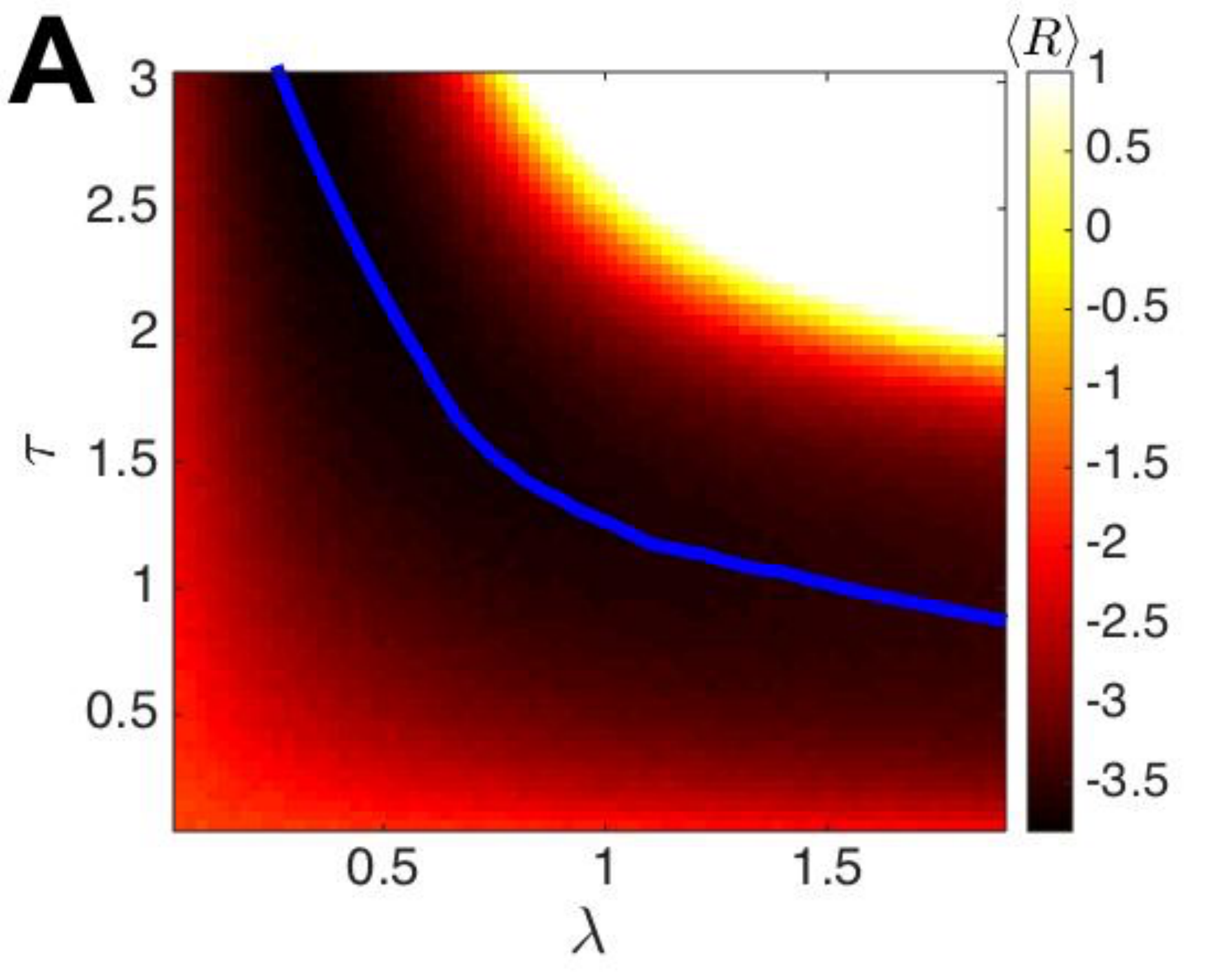}
\includegraphics[width=8cm]{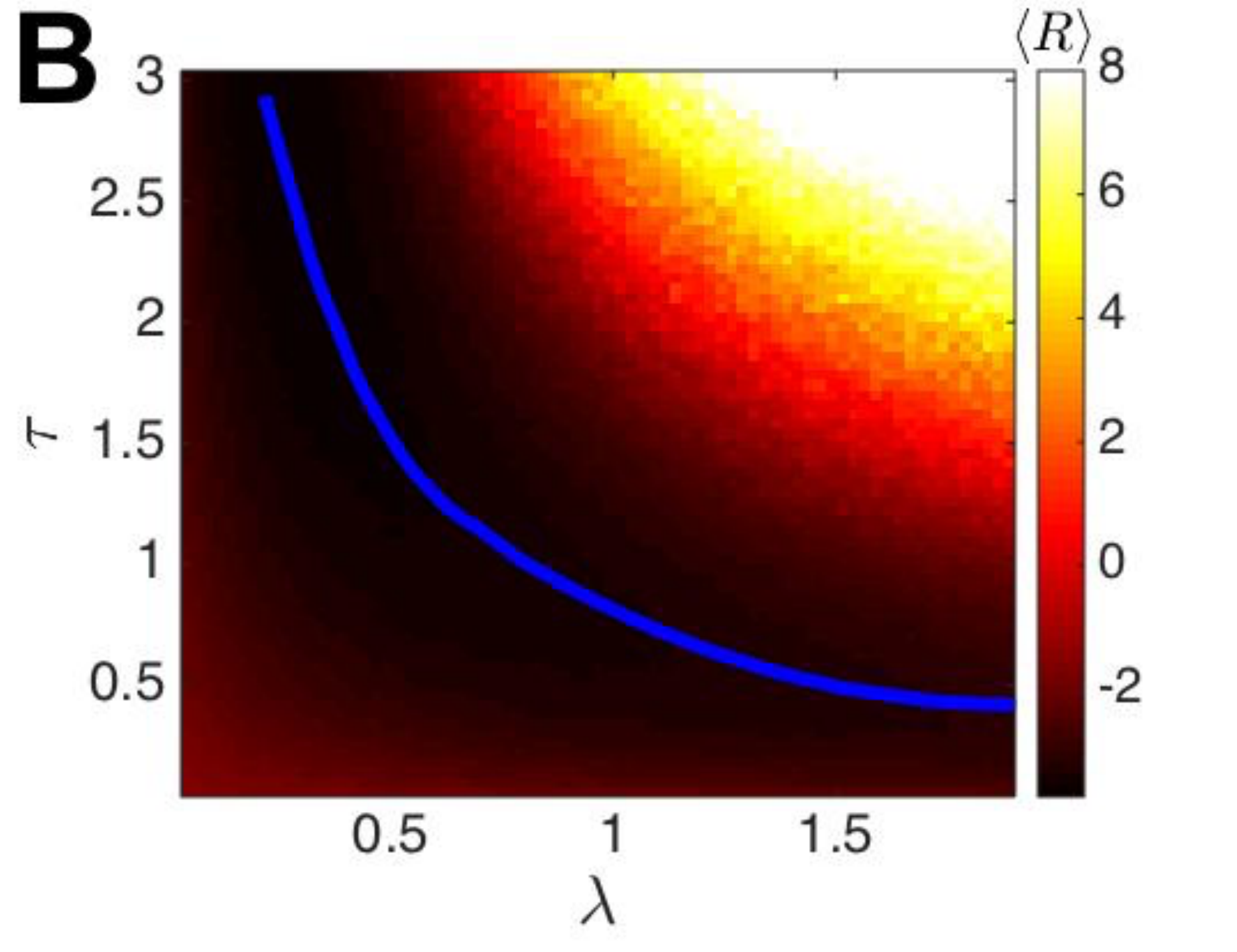}
\caption{Average log error $\langle R \rangle$ computed across realizations of (\ref{logerr}) for the discretely controlled low-dimensional approximation (\ref{noisepos}) driven by additive noise with amplitude $\epsilon = 0.1$ as described in (\ref{nfmodel}). ({\bf A}) For periodic cues with spacing $\Delta t = 4$, we find that the curve (solid line) of optimal $(\lambda, \tau)$ values has $\tau$ decreasing with $\lambda$ as in Fig. \ref{fig7}. ({\bf B}) A similar tend is observed for exponentially distributed $p(\Delta t) = \mu \e^{- \mu \Delta t}$ spacings between cue times with $\mu = 0.5$. Other parameters $v_0 = 0.15$, $\theta = 0$, and simulation time $t_f = 40$. Numerical simulations are performed using the same method as in Fig. \ref{fig7}.}
\label{fig10}
\end{figure}

We conclude by computing the average log error (\ref{logerr}) across realizations of (\ref{noisedcont}) in Fig. \ref{fig10}. Notice again that the optimal value of $\tau$, which minimizes $\langle R \rangle$ is inversely related to the strength of control. Furthermore, this trend is preserved whether cues appear periodically in time (Fig. \ref{fig10}{\bf A}) or at exponentially distributed intervals (Fig. \ref{fig10}{\bf B}).

\section{Discussion}
\label{discussion}

We have introduced and studied a neural field model of path integration with sensory feedback. Velocity-input results in the propagation of a bump attractor whose position encodes an animal's estimate of its position. Sensory information is assumed to come in the form of cues that are either constantly present, in the case of continuous feedback, or present at discrete points in time, in the case of discrete feedback. The full neural field model (\ref{nfmodel}) can be reduced to a single scalar equation (\ref{ddel2}) for the resulting position of the bump attractor. Analyzing this reduced equation, we have found that continuous control can be used to reduce the error to zero in a variety of cases. Incorporating the more realistic assumption of discrete sensory control, we find a tradeoff arises as the strength of control $\lambda$ is tuned: error reduction when cues are recent, counteracted by error increases when cues are older and irrelevant. Thus, there is an optimal control strength $\lambda$ that minimizes the long term error in the model's position estimate. This pattern holds when errors originate from spatial heterogeneities as well as dynamic fluctuations.

Our analysis has focused on one-dimensional periodic systems, wherein it is assumed the animal is navigating along a narrow annular track (Fig. \ref{fig1}{\bf B}). This was based on the protocol used in the experiments of \cite{battaglia04}, which were used to study the effect of local cues on the sharpness of neuronal place fields. However, there are several studies of navigation in two-dimensional and even three-dimensional space that demonstrate mammals' ability to use sensory cues to perform error correction \citep{geva15,solstad08}. For instance, a recent study has demonstrated that encounters with the boundaries of rectangular environments correct for the systematic drift in position representation \citep{hardcastle15}. In particular, border cells in medial entorhinal cortex (MEC) are thought to provide inputs to position-encoding grid cells when an animal senses an environmental boundary. Such recent studies are consistent with the predictions of planar models of spatial navigation based on the dynamics of velocity-driven bump attractors \citep{burak09,samsonovich97}. The model we have presented here could be extended to incorporate the effects of position-dependent cues, like boundaries, in two-dimensional domains. We expect the extension to two-dimensional neural field models should be possible through a similar negative feedback control mechanism to those presented in section \ref{model}. Our derivation of the reduced equation would then simply yield a position variable that is two-dimensional, with a correction term along each coordinate.

We also note that there is recent evidence that the position of discrete objects in the environment may be encoded by cells in the lateral entorhinal cortex (LEC) \citep{tsao13}. In particular, these cells tend to be inactive in open environments with no spatial landmarks, but they become active in the presence of objects that can help animals to orient themselves \citep{deshmukh11}. Some cells in LEC, object-trace cells, have been shown to fire when an animal encounters a location where an object previous was located, demonstrating a persistent memory of location \citep{tsao13}. If in fact such cells provide inputs to the position-encoding networks in MEC or hippocampus, LEC object cells could provide a candidate mechanism for the sensory feedback control which we have modeled here.

\section*{Acknowledgements}

This work was supported by an NSF grant (DMS-1311755). KN was supported by a Goldwater Scholarship and a University of Houston Summer Undergraduate Research Fellowship.

\bibliographystyle{spbasic} 
\bibliography{navigation}

\end{document}